\documentclass[12pt]{article}
\usepackage{amssymb}
\usepackage{amsmath}
\usepackage{graphicx}
\usepackage{indentfirst}
\usepackage{cite}
\usepackage{appendix}

\allowdisplaybreaks

\begin{document}

\title{Dissociation cross sections of $\psi (3770)$, $\psi (4040)$, 
$\psi (4160)$, and $\psi (4415)$ mesons with nucleons}
\author{Ruo-Qing Ding$^1$, Xiao-Ming Xu$^1$, and H. J. Weber$^2$}
\date{}
\maketitle \vspace{-1cm}
\centerline{$^1$Department of Physics, Shanghai University, Baoshan,
Shanghai 200444, China}
\centerline{$^2$Department of Physics, University of Virginia, Charlottesville,
VA 22904, USA}

\begin{abstract}
We study the dissociation of $\psi (3770)$, $\psi (4040)$, $\psi (4160)$, and
$\psi (4415)$ mesons in collisions with nucleons, which takes place in 
high-energy proton-nucleus collisions. 
Quark interchange between a nucleon and a $c\bar c$ meson leads to 
the dissociation of the $c\bar c$ meson. We consider the reactions:
$pR \to \Lambda_c^+ \bar{D}^0$, $pR \to \Lambda_c^+ \bar{D}^{*0}$,
$pR \to \Sigma_c^{++} D^-$, $pR \to \Sigma_c^{++} D^{*-}$,
$pR \to \Sigma_c^{+} \bar{D}^0$, $pR \to \Sigma_c^{+} \bar{D}^{*0}$,
$pR \to \Sigma_c^{*++} D^-$, $pR \to \Sigma_c^{*++} D^{*-}$,
$pR \to \Sigma_c^{*+} \bar{D}^0$, and $pR \to \Sigma_c^{*+} \bar{D}^{*0}$,
where $R$ stands for $\psi (3770)$, $\psi (4040)$, $\psi (4160)$, or 
$\psi (4415)$.
A reaction of a neutron and a $c\bar c$ meson corresponds to a reaction
of a proton and the $c\bar c$ meson by replacing the up quark with the
down quark and vice versa. Transition-amplitude formulas are derived from the
$S$-matrix element. Unpolarized cross sections are calculated with the
transition amplitudes for scattering in the prior form and in the post form.
The cross sections relate to nodes in the radial wave functions of
$\psi (3770)$, $\psi (4040)$, $\psi (4160)$, and $\psi (4415)$ mesons.
\end{abstract}



\vspace{0.5cm}
\leftline{\bf I. INTRODUCTION}
\vspace{0.5cm}

It is shown in Refs. \cite{GI,BGS,VFV} 
that $\psi (3770)$, $\psi (4040)$, $\psi (4160)$, and $\psi (4415)$ mesons
are the $1 ^3D_1$, $3 ^3S_1$, $2 ^3D_1$, and $4 ^3S_1$ states
of a charm quark and a charm antiquark. The four $c\bar c$ mesons 
have been widely studied in $e^+ e^-$ annihilation that produce hadrons
\cite{MARKI,MARKII,DASP,BES2008}, 
$\pi\pi J/\psi$ \cite{BES2005,CLEO082004,CLEO162003},
$\eta J/\psi$ \cite{CLEO082004,CLEO162003,BESIII2012,BELLE2013},
$K^+ K^- J/\psi$ \cite{CLEO162003},
$\gamma \chi_{cJ}$ $(J=1,2)$ \cite{CLEO162003,BELLE2015},
two charmed mesons \cite{BELLE2008,BABAR2009},
$D^0 D^{*-} \pi^+$ \cite{BELLE2009},
two charmed strange mesons \cite{BELLE2011,BABAR2010},
$\pi\pi h_c$ \cite{CLEO041803},
$\omega \chi_{c2}$ \cite{BESIII2016},
$\mu^+ \mu^-$ \cite{BESIII2020}, and
$\Lambda \bar{\Lambda}$ \cite{BESIII2021}.
Electron-positron annihilation produces a virtual photon which splits into a 
charm quark and a charm antiquark, and this quark-antiquark pair becomes a 
$c \bar c$ meson nonperturbatively. Production of the $\psi (3770)$
meson in $e^+ e^-$ annihilation was studied in the nonrelativistic
quantum chromodynamics (NRQCD) factorization formalism that includes
color-singlet and color-octet contributions \cite{HLC}. In Refs. 
\cite{PGK,BIO} the conversion of the photon to the $\psi (4040)$ or 
$\psi (4160)$ meson is indicated by a constant factor.

Au-Au collisions at the Relativistic Heavy Ion Collider (RHIC) and Pb-Pb
collisions at the Large Hadron Collider (LHC) produce quark-gluon plasmas.
At the critical temperature $T_{\rm c}$ the quark-gluon plasma becomes hadronic
matter. Since
$\psi (4040)$, $\psi (4160)$, and $\psi (4415)$ mesons are dissolved in 
hadronic matter when the temperature is larger than $0.97T_{\rm c}$,
$0.95T_{\rm c}$, and $0.87T_{\rm c}$, respectively \cite{wxLXW}, they
can only be produced in  hadronic matter. Therefore,
the production of $\psi (4040)$, $\psi (4160)$, and $\psi (4415)$ can be taken
as probes of hadronic matter that results from the quark-gluon plasma created 
in ultrarelativistic heavy-ion collisions. In hadronic matter
they are produced in the following reactions:
$D\bar{D} \to \rho R$, $D\bar{D}^* \to \pi R$, $D\bar{D}^* \to \rho R$, 
$D^*\bar{D}^* \to \pi R$, $D^*\bar{D}^* \to \rho R$ and so on,
where $R$ stands for $\psi (4040)$, $\psi (4160)$, or $\psi (4415)$.
Charmed mesons have been well measured in Pb-Pb collisions at the LHC.
It is shown in Ref. \cite{lyLXW} that numbers
of $\psi(4040)$, $\psi(4160)$, and $\psi(4415)$ produced in a central Pb-Pb
collision at the center-of-mass energy per nucleon-nucleon pair 
$\sqrt{s_{NN}}=5.02$ TeV are 0.25, 0.1, and 0.18, respectively. Therefore, it 
is interesting to measure $\psi (4040)$, $\psi (4160)$, and $\psi (4415)$ 
mesons produced in Pb-Pb collisions at the LHC.

Production of $D$-wave charmonia in nucleon-nucleon collisions was studied
in NRQCD in Ref. \cite{YQC}. Production cross sections depend on parton
distribution functions, short-distance processes, and nonperturbative matrix
elements of four-fermion operators. In proton-nucleus reactions a charmonium 
produced in a proton-nucleon collision further interacts with other nucleons.
The nucleon-charmonium collisions may break the charmonium, and thus reduce 
the 
charmonium number. Therefore, in the present work we study the dissociation of 
$\psi (3770)$, $\psi (4040)$, $\psi (4160)$, and $\psi (4415)$ mesons in 
collisions with nucleons. Since many experiments on $pA$ reactions have been 
carried out at the RHIC and the LHC, it is interesting to study the 
dissociation processes.

$\psi (3770)$, $\psi (4040)$, $\psi (4160)$, and $\psi (4415)$ mesons are
of special interest because they are easily produced at electron-positron
colliders. The mechanism of producing them in
proton-nucleus reactions is different from the mechanism of
producing them in electron-positron collisions. The mesons are influenced by
cold nuclear matter due to the dissociation processes and nuclear modification
of parton distribution functions. Therefore, it will be
interesting to compare the production of the mesons in proton-nucleus 
reactions with the production in electron-positron collisions in both
experiment and theory in future.

This paper is organized as follows. In Sect.~II we derive formulas of 
transition amplitudes which are used to calculate unpolarized cross sections
for dissociation of $c\bar c$ mesons in collisions with nucleons. 
In Sect.~III we present
numerical cross sections along with relevant discussions. In Sect.~IV we 
summarize the present work.

\vspace{0.5cm}
\leftline{\bf II. FORMALISM}
\vspace{0.5cm}

We consider the reaction $A+B \to C+D$ where $A$ and $C$ represent baryons and
$B$ and $D$ are mesons.
Denote by $E_{\rm i}$ and $\vec{P}_{\rm i}$ ($E_{\rm f}$ and $\vec{P}_{\rm f}$)
the total energy and the total momentum of the initial (final) baryon and the
initial (final) meson,
respectively. If $E_A$ ($E_B$, $E_C$, $E_D$) stands for the energy of 
hadron $A$ ($B$, $C$, $D$), $E_{\rm i}=E_A+E_B$ and $E_{\rm f}=E_C+E_D$. Let 
$H_I$ be the interaction potential between two constituents of hadrons in the 
reaction $A(q_1q_2q_3)+B(c\bar c) \to C(q_1q_2c)+D(q_3\bar c)$, where $q_1$,
$q_2$, and $q_3$ represent light quarks. Since the quark flavors inside baryon
$A$ differ from the charm flavor inside meson $B$, quark interchange (for
example, $q_3$ and $c$) between baryon $A$ and meson $B$ gives rise to the 
reaction. The $S$-matrix element for $A+B \to C+D$ is
\begin{equation}
S_{\rm fi} = \delta_{\rm fi} - 2\pi {\textrm i} \delta (E_{\rm f} - E_{\rm i})
<C,D \mid H_{\rm I} \mid A,B>.
\end{equation}

Let $\vec {P}_{q_1q_2q_3}$ ($\vec {P}_{q_1q_2c}^{~\prime}$) and 
$\vec {R}_{q_1q_2q_3}$ ($\vec {R}_{q_1q_2c}$) be the total momentum and the 
center-of-mass coordinate of $q_1$, $q_2$, and $q_3$ ($q_1$, $q_2$, and $c$) 
in baryon $A$ ($C$), respectively. Let $\vec {P}_{c\bar c}$
($\vec {P}_{q_3\bar c}^\prime$), $\vec {R}_{c\bar c}$
($\vec {R}_{q_3\bar c}$), and $\vec {r}_{c\bar c}$ ($\vec {r}_{q_3\bar c}$)
be the total momentum, the center-of-mass coordinate, and the relative
coordinate of $c$ and $\bar c$ ($q_3$ and $\bar c$)  of meson $B$ ($D$), 
respectively. In case that quarks $q_1$ and $q_2$ have the same mass,
we define
\begin{equation}
\vec{\rho}=\frac {1}{\sqrt 2}(\vec{r}_{q_1}-\vec{r}_{q_2}),
\end{equation}
and
\begin{equation}
\vec{\lambda}_{\rm i}=\frac {1}{\sqrt 6}(\vec{r}_{q_1}+\vec{r}_{q_2}
-2\vec{r}_{q_3}),
\end{equation}
for baryon A, and
\begin{equation}
\vec{\lambda}_{\rm f}=\frac {1}{\sqrt 6}(\vec{r}_{q_1}+\vec{r}_{q_2}
-2\vec{r}_c),
\end{equation}
for baryon $C$, where $\vec{r}_{q_1}$, $\vec{r}_{q_2}$, $\vec{r}_{q_3}$, and
$\vec{r}_c$ are the position vectors of quarks $q_1$, $q_2$, $q_3$, and 
$c$, respectively. The wave function $\mid A,B>$ of baryon $A$ and meson $B$ is
\begin{equation}
\psi_{AB}=
\frac {\textrm{e}^{\textrm{i}\vec {P}_{q_1q_2q_3}\cdot 
\vec {R}_{q_1q_2q_3}}}{\sqrt V}
\psi_{q_1q_2q_3} (\vec {\rho},\vec {\lambda}_{\rm i})
\frac {\textrm{e}^{\textrm{i}\vec {P}_{c\bar c}\cdot 
\vec {R}_{c\bar c}}}{\sqrt V}
\psi_{c\bar c} (\vec {r}_{c\bar c}),
\end{equation}
and the wave function $\mid C,D>$ of baryon $C$ and meson $D$ is
\begin{equation}
\psi_{CD}=
\frac {\textrm{e}^{\textrm{i}\vec {P}_{q_1q_2c}^\prime \cdot 
\vec {R}_{q_1q_2c}}}{\sqrt V}
\psi_{q_1q_2c} (\vec {\rho},\vec {\lambda}_{\rm f})
\frac {\textrm{e}^{\textrm{i}\vec {P}_{q_3\bar c}^\prime \cdot 
\vec {R}_{q_3\bar c}}}{\sqrt V}
\psi_{q_3\bar c} (\vec {r}_{q_3\bar c}),
\end{equation}
in which $V$ is the volume where every hadron wave function is normalized.
$\psi_{q_1q_2q_3} (\vec {\rho},\vec {\lambda}_{\rm i})$
($\psi_{q_1q_2c} (\vec {\rho},\vec {\lambda}_{\rm f})$)
is the product of the color wave function, the flavor wave function, the spin 
wave function, and the space wave function of the three quarks.
$\psi_{c\bar c} (\vec {r}_{c\bar c})$ 
($\psi_{q_3\bar c} (\vec {r}_{q_3\bar c})$)
is the product of the color wave function, the flavor wave function, the spin 
wave function, and the quark-antiquark relative-motion wave function.

With the wave functions we have
\begin{eqnarray}
& & <C,D \mid H_{\rm I} \mid A,B>   
\nonumber \\
& = & \int d^3R_{q_1q_2q_3} d^3\rho d^3\lambda_{\rm i} d^3R_{c\bar c}
d^3r_{c\bar c} \psi_{CD}^+ H_{\rm I} \psi_{AB} 
\nonumber \\
& = & \int d^3\rho d^3\lambda_{\rm i} d^3r_{c\bar c} d^3r_{q_1q_2q_3,c\bar c}
d^3R_{\rm total}
\nonumber \\
& & \frac {\psi_{q_1q_2c}^+ (\vec {\rho},\vec {\lambda}_{\rm f})}{\sqrt V}
\frac {\psi_{q_3\bar c}^+ (\vec {r}_{q_3\bar c})}{\sqrt V}
\exp(-\textrm{i}\vec{P}_{\rm f} \cdot \vec{R}_{\rm total}
-\textrm{i}\vec{p}_{q_1q_2c,q_3\bar c}^{~\prime} \cdot 
\vec{r}_{q_1q_2c,q_3\bar c})
\nonumber \\
& & H_{\rm I} 
\frac {\psi_{q_1q_2q_3} (\vec {\rho},\vec {\lambda}_{\rm i})}{\sqrt V}
\frac {\psi_{c\bar c} (\vec {r}_{c\bar c})}{\sqrt V}
\exp(\textrm{i}\vec{P}_{\rm i} \cdot 
\vec{R}_{\rm total}+\textrm{i}\vec{p}_{q_1q_2q_3,c\bar c}
\cdot \vec{r}_{q_1q_2q_3,c\bar c})
\nonumber \\
& = & 
(2\pi)^3 \delta^3 (\vec{P}_{\rm f} - \vec{P}_{\rm i})
\int d^3\rho d^3\lambda_{\rm i} d^3r_{c\bar c} d^3r_{q_1q_2q_3,c\bar c}
\frac {\psi_{q_1q_2c}^+ (\vec {\rho},\vec {\lambda}_{\rm f})}{\sqrt V}
\frac {\psi_{q_3\bar c}^+ (\vec {r}_{q_3\bar c})}{\sqrt V}
H_{\rm I} 
\nonumber \\
& &
\frac {\psi_{q_1q_2q_3} (\vec {\rho},\vec {\lambda}_{\rm i})}{\sqrt V}
\frac {\psi_{c\bar c} (\vec {r}_{c\bar c})}{\sqrt V}
\exp(-\textrm{i}\vec{p}_{q_1q_2c,q_3\bar c}^{~\prime} \cdot 
\vec{r}_{q_1q_2c,q_3\bar c}
+\textrm{i}\vec{p}_{q_1q_2q_3,c\bar c} \cdot \vec{r}_{q_1q_2q_3,c\bar c})
\nonumber \\
& = & (2\pi)^3 \delta^3 (\vec{P}_{\rm f} - \vec{P}_{\rm i})
\frac {{\cal M}_{\rm fi}}{V^2\sqrt{2E_A2E_B2E_C2E_D}},
\end{eqnarray}
where $\vec {r}_{q_1q_2q_3,c\bar c}$ ($\vec {r}_{q_1q_2c, q_3\bar c}$) and 
$\vec {p}_{q_1q_2q_3,c\bar c}$ ($\vec {p}_{q_1q_2c, q_3\bar c}^{~\prime}$) are
the relative coordinate and the relative momentum of $q_1q_2q_3$ and $c\bar c$ 
($q_1q_2c$ and $q_3\bar c$), respectively; 
$\vec{R}_{\rm total}$ is the center-of-mass coordinate of the two initial
hadrons, i.e., of the two final hadrons; $\psi_{CD}^+$ is the Hermitean 
conjugate of $\psi_{CD}$; ${\cal M}_{\rm fi}$ is the transition amplitude given
by
\begin{eqnarray}
{\cal M}_{\rm fi} & = & \sqrt{2E_A2E_B2E_C2E_D}
\int d^3\rho d^3\lambda_{\rm i} d^3r_{c\bar c} d^3r_{q_1q_2q_3,c\bar c}
\psi_{q_1q_2c}^+ (\vec {\rho},\vec {\lambda}_{\rm f})
\psi_{q_3\bar c}^+ (\vec {r}_{q_3\bar c}) H_{\rm I} 
\nonumber \\
& &
\psi_{q_1q_2q_3} (\vec {\rho},\vec {\lambda}_{\rm i})
\psi_{c\bar c} (\vec {r}_{c\bar c})
\exp(-\textrm{i}\vec{p}_{q_1q_2c,q_3\bar c}^{~\prime} \cdot 
\vec{r}_{q_1q_2c,q_3\bar c}
+\textrm{i}\vec{p}_{q_1q_2q_3,c\bar c} \cdot \vec{r}_{q_1q_2q_3,c\bar c}) .
\end{eqnarray}

The wave function of baryon $A$ and meson $B$ is
\begin{equation}
\psi_{AB} =\phi_{A\rm color} \phi_{B\rm color} \phi_{A\rm flavor} 
\phi_{B\rm flavor} \phi_{A\rm space} \chi_{S_A S_{Az}} \phi_{BJ_BJ_{Bz}},
\end{equation}
and the wave function of baryon $C$ and meson $D$ is
\begin{equation}
\psi_{CD} =\phi_{C\rm color} \phi_{D\rm color} \phi_{C\rm flavor} 
\phi_{D\rm flavor} \phi_{C\rm space} \phi_{D\rm rel} 
\chi_{S_C S_{Cz}} \chi_{S_D S_{Dz}},
\end{equation}
where $S_A$ ($S_C$, $S_D$) is the spin of hadron $A$ ($C$, $D$) with its 
magnetic projection quantum number $S_{Az}$ ($S_{Cz}$, $S_{Dz}$); 
$\phi_{A\rm color}$ ($\phi_{C\rm color}$, $\phi_{D\rm color}$), 
$\phi_{A\rm flavor}$ ($\phi_{C\rm flavor}$, $\phi_{D\rm flavor}$), and
$\chi_{S_A S_{Az}}$ ($\chi_{S_C S_{Cz}}$, $\chi_{S_D S_{Dz}}$) are the color 
wave function, the flavor wave function, and the spin wave function of hadron 
$A$ ($C$, $D$), respectively; $\phi_{A\rm space}$ ($\phi_{C\rm space}$) is
the space wave function of baryon $A$ ($C$); $\phi_{D\rm rel}$ 
is the quark-antiquark relative-motion wave function of meson $D$;
$\phi_{B\rm color}$, $\phi_{B\rm flavor}$, and $\phi_{BJ_BJ_{Bz}}$
are the color wave function, the flavor wave function,
and the space-spin wave function of meson $B$ with the 
total angular momentum $J_B$ and its $z$ component $J_{Bz}$, respectively.
Denote by $L_B$ and $S_B$ the orbital angular momentum and the spin of meson
$B$, respectively, and by $M_B$ and $S_{Bz}$ the magnetic projection
quantum numbers of $L_B$ and $S_B$. In Eq. (9)
$\phi_{BJ_BJ_{Bz}}=R_{L_B}(r_{c\bar c})$ $\sum^{L_B}_{M_B=-L_B}
\sum^{S_B}_{S_{Bz}=-S_B} (L_BM_BS_BS_{Bz} \mid J_BJ_{Bz}) Y_{L_BM_B}
\chi_{S_BS_{Bz}}$ where $R_{L_B}(r_{c\bar c})$ is the radial wave function
of the relative motion of $c$ and $\bar c$,
$(L_B M_B S_B S_{Bz} \mid J_B J_{Bz})$ are the Clebsch-Gordan coefficients,
$Y_{L_B M_B}$ are the spherical harmonics, and $\chi_{S_BS_{Bz}}$ are the 
spin wave functions.

The interaction that governs scattering in the prior form shown in Fig. 1 
is
\begin{equation}
H_{\rm I}=V_{q_1\bar c}+V_{q_2\bar c}+V_{q_3\bar c}+V_{q_1c}+V_{q_2c}+V_{q_3c},
\end{equation}
and the interaction that governs scattering in the post form shown in Fig.
2 is
\begin{equation}
H_{\rm I}=V_{q_1\bar c}+V_{q_2\bar c}+V_{c\bar c}+V_{q_1q_3}+V_{q_2q_3}
+V_{q_3c},
\end{equation}
where $V_{ab}$ is the potential between constituents $a$ and $b$.
Let $\vec{r}_{\bar c}$ be the position vector of antiquark $\bar c$. We take 
the Fourier transform of the potentials and wave functions:
\begin{equation}
V_{q_1\bar c}(\vec{r}_{q_1}-\vec{r}_{\bar c}) =
\int \frac {d^3Q}{(2\pi)^3} V_{q_1\bar c} (\vec {Q})
\textrm{e}^{\textrm{i}\vec {Q} \cdot (\vec{r}_{q_1}-\vec{r}_{\bar c})},
\end{equation}
\begin{equation}
V_{q_2\bar c}(\vec{r}_{q_2}-\vec{r}_{\bar c}) =
\int \frac {d^3Q}{(2\pi)^3} V_{q_2\bar c} (\vec {Q})
\textrm{e}^{\textrm{i}\vec {Q} \cdot (\vec{r}_{q_2}-\vec{r}_{\bar c})},
\end{equation}
\begin{equation}
V_{q_3\bar c}(\vec{r}_{q_3}-\vec{r}_{\bar c}) =
\int \frac {d^3Q}{(2\pi)^3} V_{q_3\bar c} (\vec {Q})
\textrm{e}^{\textrm{i}\vec {Q} \cdot (\vec{r}_{q_3}-\vec{r}_{\bar c})},
\end{equation}
\begin{equation}
V_{q_1c}(\vec{r}_{q_1}-\vec{r}_c) =
\int \frac {d^3Q}{(2\pi)^3} V_{q_1c} (\vec {Q})
\textrm{e}^{\textrm{i}\vec {Q} \cdot (\vec{r}_{q_1}-\vec{r}_c)},
\end{equation}
\begin{equation}
V_{q_2c}(\vec{r}_{q_2}-\vec{r}_c) =
\int \frac {d^3Q}{(2\pi)^3} V_{q_2c} (\vec {Q})
\textrm{e}^{\textrm{i}\vec {Q} \cdot (\vec{r}_{q_2}-\vec{r}_c)},
\end{equation}
\begin{equation}
V_{q_3c}(\vec{r}_{q_3}-\vec{r}_c) =
\int \frac {d^3Q}{(2\pi)^3} V_{q_3c} (\vec {Q})
\textrm{e}^{\textrm{i}\vec {Q} \cdot (\vec{r}_{q_3}-\vec{r}_c)},
\end{equation}
\begin{equation}
V_{c\bar c}(\vec{r}_c-\vec{r}_{\bar c}) =
\int \frac {d^3Q}{(2\pi)^3} V_{c\bar c} (\vec {Q})
\textrm{e}^{\textrm{i}\vec {Q} \cdot (\vec{r}_c-\vec{r}_{\bar c})},
\end{equation}
\begin{equation}
V_{q_1q_3}(\vec{r}_{q_1}-\vec{r}_{q_3}) =
\int \frac {d^3Q}{(2\pi)^3} V_{q_1q_3} (\vec {Q})
\textrm{e}^{\textrm{i}\vec {Q} \cdot (\vec{r}_{q_1}-\vec{r}_{q_3})},
\end{equation}
\begin{equation}
V_{q_2q_3}(\vec{r}_{q_2}-\vec{r}_{q_3}) =
\int \frac {d^3Q}{(2\pi)^3} V_{q_2q_3} (\vec {Q})
\textrm{e}^{\textrm{i}\vec {Q} \cdot (\vec{r}_{q_2}-\vec{r}_{q_3})},
\end{equation}
\begin{equation}
\phi_{A\rm space}(\vec{\rho},\vec{\lambda_{\rm i}}) =
\int \frac {d^3p_\rho}{(2\pi)^3}\frac {d^3p_\lambda}{(2\pi)^3} 
\phi_{A\rm space} (\vec {p}_\rho,\vec {p}_\lambda)
\textrm{e}^{\textrm{i}\vec {p}_\rho \cdot \vec{\rho} 
+\textrm{i}\vec {p}_\lambda \cdot \vec{\lambda}_{\rm i}},
\end{equation}
\begin{equation}
\phi_{BJ_BJ_{Bz}}(\vec{r}_{c\bar c}) =
\int \frac {d^3p_{c\bar c}}{(2\pi)^3} \phi_{BJ_BJ_{Bz}}
(\vec {p}_{c\bar c})
\textrm{e}^{\textrm{i}\vec {p}_{c\bar c} \cdot \vec {r}_{c\bar c}},
\end{equation}
\begin{equation}
\phi_{C\rm space}(\vec{\rho},\vec{\lambda_{\rm f}}) =
\int \frac {d^3p_\rho^\prime}{(2\pi)^3}\frac {d^3p_\lambda^\prime}{(2\pi)^3} 
\phi_{C\rm space}(\vec {p}_\rho^{~\prime},\vec {p}_\lambda^{~\prime})
\textrm{e}^{\textrm{i}\vec {p}_\rho^{~\prime} \cdot \vec{\rho} 
+\textrm{i}\vec {p}_\lambda^{~\prime} \cdot \vec{\lambda}_{\rm f}},
\end{equation}
\begin{equation}
\phi_{D\rm rel}(\vec{r}_{q_3\bar c}) =
\int \frac {d^3p_{q_3\bar c}^\prime}{(2\pi)^3} \phi_{D\rm rel}
(\vec {p}_{q_3\bar c}^{~\prime})
\textrm{e}^{\textrm{i}\vec {p}_{q_3\bar c}^{~\prime} \cdot 
\vec {r}_{q_3\bar c}},
\end{equation}
where $\vec Q$ is the momentum attached to the dot-dashed lines in Figs. 1 and
2, $\vec {p}_{c\bar c}$ is the relative 
momentum of $c$ and $\bar c$ in meson $B$, and 
$\vec {p}_{q_3\bar c}^{~\prime}$ is the relative momentum of $q_3$ and 
$\bar c$ in meson $D$. 
In momentum space the normalizations are 
\begin{equation}
\int \frac{d^3p_\rho}{(2\pi)^3} \int \frac{d^3p_\lambda}{(2\pi)^3}
\phi_{A\rm space}^+(\vec{p}_\rho,\vec{p}_\lambda)
\phi_{A\rm space}(\vec{p}_\rho,\vec{p}_\lambda)=1,
\end{equation}
\begin{equation}
\int \frac{d^3p_{c\bar c}}{(2\pi)^3}
\phi_{BJ_BJ_{Bz}}^+(\vec{p}_{c\bar c})
\phi_{BJ_BJ_{Bz}}(\vec{p}_{c\bar c})=1, 
\end{equation}
\begin{equation}
\int \frac{d^3p_\rho^\prime}{(2\pi)^3} 
\int \frac{d^3p_\lambda^\prime}{(2\pi)^3}
\phi_{C\rm space}^+(\vec{p}_\rho^{~\prime},\vec{p}_\lambda^{~\prime})
\phi_{C\rm space}(\vec{p}_\rho^{~\prime},\vec{p}_\lambda^{~\prime})=1,
\end{equation}
\begin{equation}
\int \frac{d^3p_{q_3\bar c}^\prime}{(2\pi)^3}
\phi_{D\rm rel}^+(\vec{p}_{q_3\bar c}^{~\prime})
\phi_{D\rm rel}(\vec{p}_{q_3\bar c}^{~\prime})=1.
\end{equation}
When quarks $q_1$ and $q_2$ have equal masses, their masses are indicated 
by $m$. Let $m_{q_3}$, $m_c$, and $m_{\bar c}$ stand for the $q_3$, $c$, and 
$\bar c$ masses, respectively. From Eqs. (8)-(29) we obtain
the transition amplitude for scattering in the prior form,
\begin{eqnarray}
{\cal{M}}_{\rm fi}^{\rm prior}&=&\sqrt{2E_A2E_B2E_C2E_D}
\phi^{+}_{C\rm{color}}\phi^{+}_{D\rm{color}}
\phi^{+}_{C\rm{flavor}}\phi^{+}_{D\rm{flavor}}
\chi_{S_C S_{Cz}}^+ \chi_{S_D S_{Dz}}^+
\nonumber \\
&&
\int \frac{d^3p_\rho^\prime}{(2\pi)^3}\frac{d^3p_\lambda^\prime}{(2\pi)^3}
\frac{d^3p_{q_3\bar c}^\prime}{(2\pi)^3}
\phi^{+}_{C\rm space}(\vec{p}_\rho^{~\prime},\vec{p}_\lambda^{~\prime})
\phi^{+}_{D\rm{rel}}(\vec{p}_{q_3\bar c}^{~\prime})
\nonumber \\
&&
\left\{ V_{q_1\bar c}\left( \frac{2}{\sqrt 6}\vec{p}_\lambda^{~\prime}
+\vec{p}_{q_3\bar c}^{~\prime}
-\vec{p}_{q_1q_2q_3,c\bar c}
+o_{\rm r}\vec{p}_{q_1q_2c,q_3\bar c}^{~\prime} \right) \right.
\nonumber \\
&&
\phi_{A\rm space}\left( \vec{p}_\rho^{~\prime}
-\frac{1}{\sqrt 3}\vec{p}_\lambda^{~\prime}
-\frac{\sqrt 2}{2}\vec{p}_{q_3\bar c}^{~\prime}
+\frac{\sqrt 2}{2}\vec{p}_{q_1q_2q_3,c\bar c}
-\frac {\sqrt{2}o_{\rm r}}{2}
\vec{p}_{q_1q_2c,q_3\bar c}^{~\prime}, \right.
\nonumber \\
&&
\left. -\frac{\sqrt 6}{2}\vec{p}_{q_3\bar c}^{~\prime}
+\frac{\sqrt{6}m_{q_3}}{2(2m+m_{q_3})}\vec{p}_{q_1q_2q_3,c\bar c}
+\frac {\sqrt{6}m_{q_3}}{2(m_{q_3}+m_{\bar c})}
\vec{p}_{q_1q_2c,q_3\bar c}^{~\prime} \right)
\nonumber \\
&&
\phi_{BJ_BJ_{Bz}}\left( -\frac{2}{\sqrt 6}\vec{p}_\lambda^{~\prime}
+\frac{m_c}{m_c+m_{\bar c}}\vec{p}_{q_1q_2q_3,c\bar c}
+\frac {m_c}{2m+m_c}\vec{p}_{q_1q_2c,q_3\bar c}^{~\prime} \right)
\nonumber \\
&& +V_{q_2\bar c}\left( \frac{2}{\sqrt 6}\vec{p}_\lambda^{~\prime}
+\vec{p}_{q_3\bar c}^{~\prime}
-\vec{p}_{q_1q_2q_3,c\bar c}
+o_{\rm r}\vec{p}_{q_1q_2c,q_3\bar c}^{~\prime} \right)
\nonumber \\
&&
\phi_{A\rm space} \left( \vec{p}_\rho^{~\prime}
+\frac{1}{\sqrt 3}\vec{p}_\lambda^{~\prime}
+\frac{\sqrt 2}{2}\vec{p}_{q_3\bar c}^{~\prime}
-\frac{\sqrt 2}{2}\vec{p}_{q_1q_2q_3,c\bar c}
+\frac {\sqrt{2}o_{\rm r}}{2}
\vec{p}_{q_1q_2c,q_3\bar c}^{~\prime}, \right.
\nonumber \\
&&
\left. -\frac{\sqrt 6}{2}\vec{p}_{q_3\bar c}^{~\prime}
+\frac{\sqrt{6}m_{q_3}}{2(2m+m_{q_3})}\vec{p}_{q_1q_2q_3,c\bar c}
+\frac {\sqrt{6}m_{q_3}}{2(m_{q_3}+m_{\bar c})}
\vec{p}_{q_1q_2c,q_3\bar c}^{~\prime} \right)
\nonumber \\
&&
\phi_{BJ_BJ_{Bz}}\left( -\frac{2}{\sqrt 6}\vec{p}_\lambda^{~\prime}
+\frac{m_c}{m_c+m_{\bar c}}\vec{p}_{q_1q_2q_3,c\bar c}
+\frac {m_c}{2m+m_c}\vec{p}_{q_1q_2c,q_3\bar c}^{~\prime} \right)
\nonumber\\
&& +V_{q_3\bar c} \left( \frac{2}{\sqrt 6}\vec{p}_\lambda^{~\prime}
+\vec{p}_{q_3\bar c}^{~\prime}
-\vec{p}_{q_1q_2q_3,c\bar c}
+o_{\rm r}\vec{p}_{q_1q_2c,q_3\bar c}^{~\prime} \right)
\nonumber \\
&&
\phi_{A\rm space}\left( \vec{p}_\rho^{~\prime},\vec{p}_\lambda^{~\prime}
-\frac{\sqrt{6}m}{2m+m_{q_3}}\vec{p}_{q_1q_2q_3,c\bar c}
+\frac {\sqrt{6}m}{2m+m_c}
\vec{p}_{q_1q_2c,q_3\bar c}^{~\prime} \right)
\nonumber \\
&&
\phi_{BJ_BJ_{Bz}}\left( -\frac{2}{\sqrt 6}\vec{p}_\lambda^{~\prime}
+\frac{m_c}{m_c+m_{\bar c}}\vec{p}_{q_1q_2q_3,c\bar c}
+\frac {m_c}{2m+m_c}\vec{p}_{q_1q_2c,q_3\bar c}^{~\prime} \right)
\nonumber\\
&& +V_{q_1c}\left( \frac{2}{\sqrt 6}\vec{p}_\lambda^{~\prime}
+\vec{p}_{q_3\bar c}^{~\prime}
-\vec{p}_{q_1q_2q_3,c\bar c}
+o_{\rm r}\vec{p}_{q_1q_2c,q_3\bar c}^{~\prime} \right)
\nonumber \\
&&
\phi_{A\rm space}\left( \vec{p}_\rho^{~\prime}
-\frac{1}{\sqrt 3}\vec{p}_\lambda^{~\prime}
-\frac{\sqrt 2}{2}\vec{p}_{q_3\bar c}^{~\prime}
+\frac{\sqrt 2}{2}\vec{p}_{q_1q_2q_3,c\bar c}
-\frac {\sqrt{2}o_{\rm r}}{2}
\vec{p}_{q_1q_2c,q_3\bar c}^{~\prime}, \right.
\nonumber \\
&&
\left. -\frac{\sqrt 6}{2}\vec{p}_{q_3\bar c}^{~\prime}
+\frac{\sqrt{6}m_{q_3}}{2(2m+m_{q_3})}\vec{p}_{q_1q_2q_3,c\bar c}
+\frac {\sqrt{6}m_{q_3}}{2(m_{q_3}+m_{\bar c})}
\vec{p}_{q_1q_2c,q_3\bar c}^{~\prime} \right)
\nonumber \\
&&
\phi_{BJ_BJ_{Bz}}\left( \vec{p}_{q_3\bar c}^{~\prime}
-\frac{m_{\bar c}}{m_c+m_{\bar c}}\vec{p}_{q_1q_2q_3,c\bar c}
+\frac {m_{\bar c}}{m_{q_3}+m_{\bar c}}\vec{p}_{q_1q_2c,q_3\bar c}^{~\prime}
\right)
\nonumber\\
&& +V_{q_2c}\left( \frac{2}{\sqrt 6}\vec{p}_\lambda^{~\prime}
+\vec{p}_{q_3\bar c}^{~\prime}
-\vec{p}_{q_1q_2q_3,c\bar c}
+o_{\rm r}\vec{p}_{q_1q_2c,q_3\bar c}^{~\prime} \right)
\nonumber \\
&&
\phi_{A\rm space}\left( \vec{p}_\rho^{~\prime}
+\frac{1}{\sqrt 3}\vec{p}_\lambda^{~\prime}
+\frac{\sqrt 2}{2}\vec{p}_{q_3\bar c}^{~\prime}
-\frac{\sqrt 2}{2}\vec{p}_{q_1q_2q_3,c\bar c}
+\frac {\sqrt{2}o_{\rm r}}{2}
\vec{p}_{q_1q_2c,q_3\bar c}^{~\prime}, \right.
\nonumber \\
&&
\left. -\frac{\sqrt 6}{2}\vec{p}_{q_3\bar c}^{~\prime}
+\frac{\sqrt{6}m_{q_3}}{2(2m+m_{q_3})}\vec{p}_{q_1q_2q_3,c\bar c}
+\frac {\sqrt{6}m_{q_3}}{2(m_{q_3}+m_{\bar c})}
\vec{p}_{q_1q_2c,q_3\bar c}^{~\prime} \right)
\nonumber \\
&&
\phi_{BJ_BJ_{Bz}}\left( \vec{p}_{q_3\bar c}^{~\prime}
-\frac{m_{\bar c}}{m_c+m_{\bar c}}\vec{p}_{q_1q_2q_3,c\bar c}
+\frac {m_{\bar c}}{m_{q_3}+m_{\bar c}}\vec{p}_{q_1q_2c,q_3\bar c}^{~\prime}
\right)
\nonumber\\
&& +V_{q_3c}\left( \frac{2}{\sqrt 6}\vec{p}_\lambda^{~\prime}
+\vec{p}_{q_3\bar c}^{~\prime}
-\vec{p}_{q_1q_2q_3,c\bar c}
+o_{\rm r}\vec{p}_{q_1q_2c,q_3\bar c}^{~\prime} \right)
\nonumber \\
&&
\phi_{A\rm space}\left( \vec{p}_\rho^{~\prime},\vec{p}_\lambda^{~\prime}
-\frac{\sqrt{6}m}{2m+m_{q_3}}\vec{p}_{q_1q_2q_3,c\bar c}
+\frac {\sqrt{6}m}{2m+m_c}
\vec{p}_{q_1q_2c,q_3\bar c}^{~\prime} \right)
\nonumber \\
&&
\phi_{BJ_BJ_{Bz}}\left. \left( \vec{p}_{q_3\bar c}^{~\prime}
-\frac{m_{\bar c}}{m_c+m_{\bar c}}\vec{p}_{q_1q_2q_3,c\bar c}
+\frac {m_{\bar c}}{m_{q_3}+m_{\bar c}}\vec{p}_{q_1q_2c,q_3\bar c}^{~\prime}
\right) \right\}
\nonumber\\
&&
\phi_{A\rm{color}}\phi_{B\rm{color}}\phi_{A\rm{flavor}}\phi_{B\rm{flavor}}
\chi_{S_A S_{Az}},
\end{eqnarray}
with $o_{\rm r}=(2mm_{\bar c}-m_{q_3}m_c)/[(2m+m_c)(m_{q_3}+m_{\bar c})]$,
and the transition amplitude for scattering in the post form,
\begin{eqnarray}
{\cal{M}}_{\rm fi}^{\rm post}&=&\sqrt{2E_A2E_B2E_C2E_D}
\phi^{+}_{C\rm{color}}\phi^{+}_{D\rm{color}}
\phi^{+}_{C\rm{flavor}}\phi^{+}_{D\rm{flavor}}
\chi_{S_C S_{Cz}}^+ \chi_{S_D S_{Dz}}^+
\nonumber\\
&&
\int \frac{d^3p_\rho^\prime}{(2\pi)^3}\frac{d^3p_\lambda^\prime}{(2\pi)^3}
\frac{d^3p_{q_3\bar c}^\prime}{(2\pi)^3}
\phi^{+}_{C\rm space}(\vec{p}_\rho^{~\prime},\vec{p}_\lambda^{~\prime})
\phi^{+}_{D\rm{rel}}(\vec{p}_{q_3\bar c}^{~\prime})
\nonumber\\
&&
\left\{ V_{q_1\bar c}\left( \frac{2}{\sqrt 6}\vec{p}_\lambda^{~\prime}
+\vec{p}_{q_3\bar c}^{~\prime}
-\vec{p}_{q_1q_2q_3,c\bar c}
+o_{\rm r}\vec{p}_{q_1q_2c,q_3\bar c}^{~\prime} \right) \right.
\nonumber \\
&&
\phi_{A\rm space}\left( \vec{p}_\rho^{~\prime}
-\frac{1}{\sqrt 3}\vec{p}_\lambda^{~\prime}
-\frac{\sqrt 2}{2}\vec{p}_{q_3\bar c}^{~\prime}
+\frac{\sqrt 2}{2}\vec{p}_{q_1q_2q_3,c\bar c}
-\frac {\sqrt{2}o_{\rm r}}{2}
\vec{p}_{q_1q_2c,q_3\bar c}^{~\prime}, \right.
\nonumber \\
&&
\left. -\frac{\sqrt 6}{2}\vec{p}_{q_3\bar c}^{~\prime}
+\frac{\sqrt{6}m_{q_3}}{2(2m+m_{q_3})}\vec{p}_{q_1q_2q_3,c\bar c}
+\frac {\sqrt{6}m_{q_3}}{2(m_{q_3}+m_{\bar c})}
\vec{p}_{q_1q_2c,q_3\bar c}^{~\prime} \right)
\nonumber \\
&&
\phi_{BJ_BJ_{Bz}}\left( -\frac{2}{\sqrt 6}\vec{p}_\lambda^{~\prime}
+\frac{m_c}{m_c+m_{\bar c}}\vec{p}_{q_1q_2q_3,c\bar c}
+\frac {m_c}{2m+m_c}\vec{p}_{q_1q_2c,q_3\bar c}^{~\prime} \right)
\nonumber \\
&& +V_{q_2\bar c}\left( \frac{2}{\sqrt 6}\vec{p}_\lambda^{~\prime}
+\vec{p}_{q_3\bar c}^{~\prime}
-\vec{p}_{q_1q_2q_3,c\bar c}
+o_{\rm r}\vec{p}_{q_1q_2c,q_3\bar c}^{~\prime} \right)
\nonumber \\
&&
\phi_{A\rm space}\left( \vec{p}_\rho^{~\prime}
+\frac{1}{\sqrt 3}\vec{p}_\lambda^{~\prime}
+\frac{\sqrt 2}{2}\vec{p}_{q_3\bar c}^{~\prime}
-\frac{\sqrt 2}{2}\vec{p}_{q_1q_2q_3,c\bar c}
+\frac {\sqrt{2}o_{\rm r}}{2}
\vec{p}_{q_1q_2c,q_3\bar c}^{~\prime}, \right.
\nonumber \\
&&
\left. -\frac{\sqrt 6}{2}\vec{p}_{q_3\bar c}^{~\prime}
+\frac{\sqrt{6}m_{q_3}}{2(2m+m_{q_3})}\vec{p}_{q_1q_2q_3,c\bar c}
+\frac {\sqrt{6}m_{q_3}}{2(m_{q_3}+m_{\bar c})}
\vec{p}_{q_1q_2c,q_3\bar c}^{~\prime} \right)
\nonumber \\
&&
\phi_{BJ_BJ_{Bz}}\left( -\frac{2}{\sqrt 6}\vec{p}_\lambda^{~\prime}
+\frac{m_c}{m_c+m_{\bar c}}\vec{p}_{q_1q_2q_3,c\bar c}
+\frac {m_c}{2m+m_c}\vec{p}_{q_1q_2c,q_3\bar c}^{~\prime} \right)
\nonumber \\
&& +V_{q_3c}\left( \frac{2}{\sqrt 6}\vec{p}_\lambda^{~\prime}
+\vec{p}_{q_3\bar c}^{~\prime}
-\vec{p}_{q_1q_2q_3,c\bar c}
+o_{\rm r}\vec{p}_{q_1q_2c,q_3\bar c}^{~\prime} \right)
\nonumber \\
&&
\phi_{A\rm space}\left( \vec{p}_\rho^{~\prime},\vec{p}_\lambda^{~\prime}
-\frac{\sqrt{6}m}{2m+m_{q_3}}\vec{p}_{q_1q_2q_3,c\bar c}
+\frac {\sqrt{6}m}{2m+m_c}
\vec{p}_{q_1q_2c,q_3\bar c}^{~\prime} \right)
\nonumber \\
&&
\phi_{BJ_BJ_{Bz}}\left. \left( \vec{p}_{q_3\bar c}^{~\prime}
-\frac{m_{\bar c}}{m_c+m_{\bar c}}\vec{p}_{q_1q_2q_3,c\bar c}
+\frac {m_{\bar c}}{m_{q_3}+m_{\bar c}}\vec{p}_{q_1q_2c,q_3\bar c}^{~\prime}
\right) \right\}
\nonumber\\
&&
\phi_{A\rm{color}}\phi_{B\rm{color}}\phi_{A\rm{flavor}}\phi_{B\rm{flavor}}
\chi_{S_A S_{Az}}
\nonumber \\
&& +\sqrt{2E_A2E_B2E_C2E_D}
\phi^{+}_{C\rm{color}}\phi^{+}_{D\rm{color}}
\phi^{+}_{C\rm{flavor}}\phi^{+}_{D\rm{flavor}}
\chi_{S_C S_{Cz}}^+ \chi_{S_D S_{Dz}}^+
\nonumber \\
&&
\int \frac{d^3p_\rho}{(2\pi)^3}\frac{d^3p_\lambda}{(2\pi)^3}
\frac{d^3p_{c\bar c}}{(2\pi)^3}
\nonumber \\
&&
\left\{ \phi^{+}_{C\rm space}\left( \vec{p}_\rho,\vec{p}_\lambda
+\frac{\sqrt{6}m}{2m+m_{q_3}}\vec{p}_{q_1q_2q_3,c\bar c}
-\frac{\sqrt{6}m}{2m+m_c}\vec{p}_{q_1q_2c,q_3\bar c}^{~\prime} \right) \right.
\nonumber \\
&&
\phi^{+}_{D\rm{rel}}\left( -\frac{2}{\sqrt 6}\vec{p}_\lambda
+\frac{m_{q_3}}{2m+m_{q_3}}\vec{p}_{q_1q_2q_3,c\bar c}
+\frac {m_{q_3}}{m_{q_3}+m_{\bar c}}
\vec{p}_{q_1q_2c,q_3\bar c}^{~\prime} \right)
\nonumber \\
&&
V_{c\bar c}\left( -\frac{2}{\sqrt 6}\vec{p}_\lambda
-\vec{p}_{c\bar c}
-o_{\rm t}\vec{p}_{q_1q_2q_3,c\bar c}
+\vec{p}_{q_1q_2c,q_3\bar c}^{~\prime} \right)
\nonumber \\
&& +\phi^{+}_{C\rm space}\left( \vec{p}_\rho
-\frac{1}{\sqrt 3}\vec{p}_\lambda
-\frac {\sqrt 2}{2}\vec{p}_{c\bar c}
-\frac{\sqrt{2}o_{\rm t}}{2}
\vec{p}_{q_1q_2q_3,c\bar c}
+\frac {\sqrt 2}{2}\vec{p}_{q_1q_2c,q_3\bar c}^{~\prime}, \right.
\nonumber \\
&&
\left. -\frac {\sqrt 6}{2}\vec{p}_{c\bar c}
+\frac{\sqrt{6}m_c}{2(m_c+m_{\bar c})}\vec{p}_{q_1q_2q_3,c\bar c}
+\frac{\sqrt{6}m_c}{2(2m+m_c)}\vec{p}_{q_1q_2c,q_3\bar c}^{~\prime} \right)
\nonumber \\
&&
\phi^{+}_{D\rm{rel}}\left( \vec{p}_{c\bar c}
+\frac{m_{\bar c}}{m_c+m_{\bar c}}\vec{p}_{q_1q_2q_3,c\bar c}
-\frac {m_{\bar c}}{m_{q_3}+m_{\bar c}}
\vec{p}_{q_1q_2c,q_3\bar c}^{~\prime} \right)
\nonumber \\
&&
V_{q_1q_3}\left( -\frac{2}{\sqrt 6}\vec{p}_\lambda
-\vec{p}_{c\bar c}
-o_{\rm t}\vec{p}_{q_1q_2q_3,c\bar c}
+\vec{p}_{q_1q_2c,q_3\bar c}^{~\prime} \right)
\nonumber \\
&& +\phi^{+}_{C\rm space}\left( \vec{p}_\rho
+\frac{1}{\sqrt 3}\vec{p}_\lambda
+\frac {\sqrt 2}{2}\vec{p}_{c\bar c}
+\frac{\sqrt{2}o_{\rm t}}{2}
\vec{p}_{q_1q_2q_3,c\bar c}
-\frac {\sqrt 2}{2}\vec{p}_{q_1q_2c,q_3\bar c}^{~\prime}, \right.
\nonumber \\
&&
\left. -\frac {\sqrt 6}{2}\vec{p}_{c\bar c}
+\frac{\sqrt{6}m_c}{2(m_c+m_{\bar c})}\vec{p}_{q_1q_2q_3,c\bar c}
+\frac{\sqrt{6}m_c}{2(2m+m_c)}\vec{p}_{q_1q_2c,q_3\bar c}^{~\prime} \right)
\nonumber \\
&&
\phi^{+}_{D\rm{rel}}\left( \vec{p}_{c\bar c}
+\frac{m_{\bar c}}{m_c+m_{\bar c}}\vec{p}_{q_1q_2q_3,c\bar c}
-\frac {m_{\bar c}}{m_{q_3}+m_{\bar c}}
\vec{p}_{q_1q_2c,q_3\bar c}^{~\prime} \right)
\nonumber \\
&&
V_{q_2q_3}\left. \left( -\frac{2}{\sqrt 6}\vec{p}_\lambda
-\vec{p}_{c\bar c}
-o_{\rm t}\vec{p}_{q_1q_2q_3,c\bar c}
+\vec{p}_{q_1q_2c,q_3\bar c}^{~\prime} \right) \right\}
\nonumber \\
&&
\phi_{A\rm space}(\vec{p}_\rho,\vec{p}_\lambda)
\phi_{BJ_BJ_{Bz}}(\vec{p}_{c\bar c})
\phi_{A\rm{color}}\phi_{B\rm{color}}\phi_{A\rm{flavor}}\phi_{B\rm{flavor}}
\chi_{S_A S_{Az}}.
\end{eqnarray}
with $o_{\rm t}=(2mm_{\bar c}-m_{q_3}m_c)/[(m_c+m_{\bar c})(2m+m_{q_3})]$.
The variables $\vec{p}_\rho$ and $\vec{p}_\lambda$ in 
$\phi_{A\rm space}(\vec{p}_\rho,$ $\vec{p}_\lambda)$, 
$\vec{p}_{c\bar c}$ in $\phi_{BJ_BJ_{Bz}}(\vec{p}_{c\bar c})$,
$\vec{p}_\rho^{~\prime}$ and $\vec{p}_\lambda^{~\prime}$ in 
$\phi_{C\rm space}(\vec{p}_\rho^{~\prime}, \vec{p}_\lambda^{~\prime})$, and
$\vec{p}_{q_3\bar c}^{~\prime}$ in 
$\phi_{D\rm rel}(\vec{p}_{q_3\bar c}^{~\prime})$ equal the expressions 
enclosed by the parentheses that follow $\phi_{A\rm space}$, 
$\phi_{BJ_BJ_{Bz}}$, $\phi_{C\rm space}$, and $\phi_{D\rm rel}$.

With the transition amplitudes the unpolarized cross section for $A+B \to C+D$
is
\begin{eqnarray}
\sigma^{\rm unpol}(\sqrt {s}) & = & \frac {1}{(2J_A+1)(2J_B+1)}
\frac{1}{64\pi s}\frac{|\vec{P}^{\prime }(\sqrt{s})|}{|\vec{P}(\sqrt{s})|}
              \nonumber    \\
& & \int_{0}^{\pi }d\theta \sum\limits_{J_{Az}J_{Bz}J_{Cz}J_{Dz}}
(\mid {\cal M}_{\rm fi}^{\rm prior} \mid^2 
+ \mid {\cal M}_{\rm fi}^{\rm post} \mid^2) \sin \theta ,
\end{eqnarray}
where $s$ is the Mandelstam variable obtained from the four-momenta $P_A$ and
$P_B$ of hadrons $A$ and $B$ by $s=(P_A+P_B)^2$; $J_A$ ($J_B$, $J_C$, $J_D$) 
and $J_{Az}$ ($J_{Bz}$, $J_{Cz}$, $J_{Dz}$) of hadron $A$ ($B$, $C$, $D$) are 
the total angular momentum and its $z$ component, respectively; $\theta$ is
the angle between $\vec P$ and $\vec{P}^\prime$ which are the three-dimensional
momentum components of baryons $A$ and $C$ in the center-of-momentum frame of 
the initial baryon and the initial meson, respectively.
We calculate the cross section in the center-of-momentum frame.

\vspace{0.5cm}
\leftline{\bf III. NUMERICAL CROSS SECTIONS AND DISCUSSIONS }
\vspace{0.5cm}

We use the notation
$ D= \left( \begin{array}{c} D^+ \\ D^0 \end{array} \right) $,
$\bar{D}= \left( \begin{array}{c} \bar{D}^0 \\ D^- \end{array} \right)$,
$ D^*= \left( \begin{array}{c} D^{*+} \\ D^{*0} \end{array} \right)$, and
$\bar{D}^*= \left( \begin{array}{c} \bar{D}^{*0} \\ D^{*-} \end{array}
\right)$.
We consider the following reactions:
\begin{displaymath}
pR \to \Lambda_c^+ \bar{D}^0,~~~~~~pR \to \Lambda_c^+ \bar{D}^{*0},
\end{displaymath}
\begin{displaymath}
pR \to \Sigma_c^{++} D^-,~~~~~~pR \to \Sigma_c^{++} D^{*-},
\end{displaymath}
\begin{displaymath}
pR \to \Sigma_c^{+} \bar{D}^0,~~~~~~pR \to \Sigma_c^{+} \bar{D}^{*0},
\end{displaymath}
\begin{displaymath}
pR \to \Sigma_c^{*++} D^-,~~~~~~pR \to \Sigma_c^{*++} D^{*-},
\end{displaymath}
\begin{displaymath}
pR \to \Sigma_c^{*+} \bar{D}^0,~~~~~~pR \to \Sigma_c^{*+} \bar{D}^{*0},
\end{displaymath}
where $R$ stands for $\psi (3770)$, $\psi (4040)$, $\psi (4160)$, or
$\psi (4415)$. By replacing the up quark with the down quark and vice versa
in these ten reactions, they give ten reactions of a neutron
and a $c\bar c$ meson. Since the cross section for
$nR \to \Lambda_c^+ D^-$ ($nR \to \Lambda_c^+ D^{*-}$, 
$nR \to \Sigma_c^0 \bar{D}^0$, $nR \to \Sigma_c^0 \bar{D}^{*0}$,
$nR \to \Sigma_c^+ D^-$, $nR \to \Sigma_c^+ D^{*-}$, 
$nR \to \Sigma_c^{*0} \bar{D}^0$, $nR \to \Sigma_c^{*0} \bar{D}^{*0}$,
$nR \to \Sigma_c^{*+} D^-$, $nR \to \Sigma_c^{*+} D^{*-}$) equals the one for
$pR \to \Lambda_c^+ \bar{D}^0$ ($pR \to \Lambda_c^+ \bar{D}^{*0}$,
$pR \to \Sigma_c^{++} D^-$, $pR \to \Sigma_c^{++} D^{*-}$,
$pR \to \Sigma_c^{+} \bar{D}^0$, $pR \to \Sigma_c^{+} \bar{D}^{*0}$,
$pR \to \Sigma_c^{*++} D^-$, $pR \to \Sigma_c^{*++} D^{*-}$,
$pR \to \Sigma_c^{*+} \bar{D}^0$, $pR \to \Sigma_c^{*+} \bar{D}^{*0}$), 
it is enough to only discuss reactions of the proton and the $c\bar c$
meson in this section. 
We calculate unpolarized cross sections for these reactions
with Eq. (32). As seen in Eqs. (30) and (31), ${\cal{M}}_{\rm fi}^{\rm prior}$
and ${\cal{M}}_{\rm fi}^{\rm post}$ used in Eq. (32) involve 
$\phi_{BJ_BJ_{Bz}}$ and $\phi_{D\rm rel}$. The two wave functions
are obtained from solutions of the Schr\"odinger equation with the
potential between constituents $a$ and $b$ in coordinate space,
\begin{eqnarray}
V_{ab}(\vec{r}_{ab}) & = &
- \frac {\vec{\lambda}_a}{2} \cdot \frac {\vec{\lambda}_b}{2} \frac {3}{4} k
r_{ab}
+ \frac {\vec{\lambda}_a}{2} \cdot \frac {\vec{\lambda}_b}{2}
\frac {6\pi}{25} \frac {v(\lambda r_{ab})}{r_{ab}}
\nonumber  \\
& & -\frac {\vec{\lambda}_a}{2} \cdot \frac {\vec{\lambda}_b}{2}
\frac {16\pi^2}{25}\frac{d^3}{\pi^{3/2}}\exp(-d^2r^2_{ab}) 
\frac {\vec {s}_a \cdot \vec {s}_b} {m_am_b}
+\frac {\vec{\lambda}_a}{2} \cdot \frac {\vec{\lambda}_b}{2}\frac {4\pi}{25}
\frac {1} {r_{ab}} \frac {d^2v(\lambda r_{ab})}{dr_{ab}^2} 
\frac {\vec {s}_a \cdot \vec {s}_b}{m_am_b}
\nonumber  \\
& & -\frac {\vec{\lambda}_a}{2} \cdot \frac {\vec{\lambda}_b}{2}
\frac {6\pi}{25m_am_b}\left[ v(\lambda r_{ab}) 
-r_{ab}\frac {dv(\lambda r_{ab})}{dr_{ab}} +\frac{r_{ab}^2}{3}
\frac {d^2v(\lambda r_{ab})}{dr_{ab}^2} \right]
\nonumber  \\
& & \left( \frac{3\vec {s}_a \cdot \vec{r}_{ab}\vec {s}_b \cdot \vec{r}_{ab}}
{r_{ab}^5} -\frac {\vec {s}_a \cdot \vec {s}_b}{r_{ab}^3} \right) ,
\end{eqnarray}
where $\vec {r}_{ab}$ is the relative coordinate of constituents $a$ and $b$;
$k=0.153$ GeV$^2$ and $\lambda=0.39$ GeV; $m_a$, $\vec {s}_a$, and 
$\vec{\lambda}_a$ are individually
the mass, the spin, and the Gell-Mann matrices for the
color generators of constituent $a$; the function $v$ is given by 
Buchm\"uller and Tye in Ref. \cite{BT}; the quantity $d$ is 
\begin{eqnarray}
d^2=d_{\alpha}^2\left[\frac{1}{2}+\frac{1}{2}\left(\frac{4m_a m_b}{(m_a+m_b)^2}
\right)^4\right]+d_{\beta}^2\left(\frac{2m_am_b}{m_a+m_b}\right)^2,
\end{eqnarray}
where $d_{\alpha}=0.34$ GeV and $d_{\beta}=0.45$.
The potential originates from quantum chromodynamics (QCD) \cite{BT}. The first
two terms are the Buchm\"uller-Tye potential, and the other
terms come from one-gluon exchange plus perturbative one- and two-loop
corrections \cite{Xu2002}. 

The function $v(x)$ manifests one-gluon exchange plus perturbative one- and 
two-loop corrections between constituents $a$ and $b$. It 
increases from 0 to 1 when $x$ increases from 0 to the positive infinity.
Consequently, the second term is not a color Coulomb potential.

One-gluon exchange between two constituents gives rise to the Fermi contact 
term $-\frac {\vec{\lambda}_a}{2} \cdot \frac {\vec{\lambda}_b}{2}
\frac {16\pi^2}{25}\delta^3 (\vec{r}_{ab})
\frac {\vec {s}_a \cdot \vec {s}_b} {m_am_b}$. The $\delta^3 (\vec{r}_{ab})$
function fixes the positions of the two constituents to $\vec{r}_{ab}=0$.
However, the constituent positions fluctuate in the presence of one- and 
two-loop
corrections. To allow the position fluctuation, $\delta^3 (\vec{r}_{ab})$ is
replaced with $\frac{d^3}{\pi^{3/2}}\exp(-d^2r^2_{ab})$ so as to arrive at
the third term on the right-hand side of Eq. (33), which is the smearing
of the  Fermi contact term \cite{GI}. 
The Gaussian has a width of $2\sqrt{\ln 2}/d$, and
$d^{-1}$ indicates the fluctuation size. The larger is $d$, the smaller is
the fluctuation size. $d$ depends on constituent masses. When $m_a=m_b$,
$d^2=d_\alpha^2+d_\beta^2 m_b^2$. When $m_a \gg m_b$, 
$d^2 \approx d_\alpha^2/2+4d_\beta^2 m_b^2$. In the two cases the $d_\alpha$ 
term gives a constant value to $d$, and the $d_\beta$ term is proportional to
$m_b^2$. The two terms provide different mass dependence. Since 
$d^2 > d_\alpha^2/2$, the parameter $d_\alpha$ reflects the fact that
in a confined system the smearing must be limited. 

The masses of the up quark, the
down quark, the strange quark, and the charm quark are 0.32 GeV, 0.32 GeV,
0.5 GeV, and 1.51 GeV, respectively. Solving the Schr\"odinger equation with
$V_{ab}$, we obtain meson masses that are close to
the experimental masses of $\pi$, $\rho$, $K$, $K^*$, $D$, $D^*$,
$D_s$, $D^*_s$, $J/\psi$, $\chi_{c}$, $\psi'$, $\psi (3770)$, $\psi (4040)$, 
$\psi (4160)$, and $\psi (4415)$ 
mesons listed in Ref.~\cite{PDG}. The experimental data of $S$-wave $I=2$
elastic phase shifts for $\pi \pi$ scattering \cite{pipiqi} are 
reproduced in the Born approximation.

${\cal{M}}_{\rm fi}^{\rm prior}$ and ${\cal{M}}_{\rm fi}^{\rm post}$
involve the space wave functions $\phi_{A\rm space}$ and $\phi_{C\rm space}$.
The space wave functions of ground-state baryons are usually assumed to be 
harmonic-oscillator wave functions \cite{CI,HBBS}:
\begin{equation}
\phi_{A\rm space}(\vec{\rho}, \vec{\lambda}_{\rm i})=
\left(\frac {\alpha_\rho \alpha_{\lambda_{\rm i}}}{\pi}\right)^{1.5}
\exp \left(-\frac{\alpha_\rho^2\vec{\rho}^{~2}
+\alpha_{\lambda_{\rm i}}^2{\vec{\lambda}_{\rm i}}^2}{2}\right) ,
\end{equation}
and $\phi_{C\rm space}(\vec{\rho}, \vec{\lambda}_{\rm f})$ is obtained from
$\phi_{A\rm space}(\vec{\rho}, \vec{\lambda}_{\rm i})$ by replacing
$\lambda_{\rm i}$ with $\lambda_{\rm f}$.
The wave function $\psi_{q_1q_2q_3} (\vec {\rho},\vec {\lambda}_{\rm i})$ in 
Eq. (5) is
\begin{equation}
\psi_{q_1q_2q_3} (\vec {\rho},\vec {\lambda}_{\rm i})
=\phi_{A\rm color} \phi_{A\rm flavor} 
\phi_{A\rm space}(\vec{\rho}, \vec{\lambda}_{\rm i}) \chi_{S_A S_{Az}},
\end{equation}
and $\psi_{q_1q_2c}$ in Eq. (6) is given from $\psi_{q_1q_2q_3}$ by 
replacing $q_3$ ($\vec{\lambda}_{\rm i}$, $A$) with $c$ 
($\vec{\lambda}_{\rm f}$, $C$).
Masses of baryons in the baryon octet and the baryon decuplet are given by
\begin{eqnarray}
m_{\rm B} & = & 2m + m_{q_3} + \int d^3\rho d^3\lambda_{\rm i}
\psi_{q_1q_2q_3}^+(\vec{\rho}, \vec{\lambda}_{\rm i})
\left[ \frac{\vec{\nabla}^2_{\vec{\rho}}}{2m}
+ \frac{\vec{\nabla}^2_{\vec{\lambda}_{\rm i}}}{2m_{\lambda_{\rm i}}} \right.
\nonumber \\
& & \left. +V_{q_1q_2} (\vec{r}_{q_1q_2})+V_{q_2q_3} (\vec{r}_{q_2q_3})
+V_{q_3q_1} (\vec{r}_{q_3q_1}) \right]
\psi_{q_1q_2q_3}(\vec{\rho}, \vec{\lambda}_{\rm i}),
\end{eqnarray}
with $m_{\lambda_{\rm i}}=\frac{3mm_{q_3}}{2m+m_{q_3}}$.
Replacing $q_3$ ($\vec{\lambda}_{\rm i}$) with $c$ ($\vec{\lambda}_{\rm f}$),
Eq. (37) is used to calculate
masses of ground-state charmed baryons.
Let $m_p$, $m_{\Lambda_c^+}$, $m_{\Sigma_c^{++}}$, $m_{\Sigma_c^{+}}$,
$m_{\Sigma_c^{*++}}$, and $m_{\Sigma_c^{*+}}$ represent the experimental 
masses of $p$, $\Lambda_c^+$, $\Sigma_c^{++}$, 
$\Sigma_c^+$, $\Sigma_c^{*++}$, and $\Sigma_c^{*+}$ baryons, respectively.
Fits to the experimental masses of the six baryons give
\begin{displaymath}
\alpha_\rho=0.3~{\rm GeV},~~~~~~\alpha_{\lambda_{\rm i}}=0.3~{\rm GeV},
~~~~~~{m_p=0.938272~{\rm GeV};}
\end{displaymath}
\begin{displaymath}
\alpha_\rho=0.222594~{\rm GeV},~~~~~~\alpha_{\lambda_{\rm f}}=0.43~{\rm GeV},
~~~~~~{m_{\Lambda_c^+}=2.28646~{\rm GeV};}
\end{displaymath}
\begin{displaymath}
\alpha_\rho=0.196273~{\rm GeV},~~~~~~\alpha_{\lambda_{\rm f}}=0.43~{\rm GeV},
~~~~~~{m_{\Sigma_c^{++}}=2.45397~{\rm GeV};}
\end{displaymath}
\begin{displaymath}
\alpha_\rho=0.19642~{\rm GeV},~~~~~~\alpha_{\lambda_{\rm f}}=0.43~{\rm GeV},
~~~~~~{m_{\Sigma_c^{+}}=2.4529~{\rm GeV};}
\end{displaymath}
\begin{displaymath}
\alpha_\rho=0.19092~{\rm GeV},~~~~~~\alpha_{\lambda_{\rm f}}=0.43~{\rm GeV},
~~~~~~{m_{\Sigma_c^{*++}}=2.51841~{\rm GeV};}
\end{displaymath}
\begin{displaymath}
\alpha_\rho=0.19105~{\rm GeV},~~~~~~\alpha_{\lambda_{\rm f}}=0.43~{\rm GeV},
~~~~~~{m_{\Sigma_c^{*+}}=2.5175~{\rm GeV}.}
\end{displaymath}

Using the mesonic quark-antiquark relative-motion wave functions and the space
wave functions of baryons, we obtain unpolarized cross sections for 
dissociation of $\psi (3770)$, $\psi (4040)$, $\psi (4160)$, and $\psi (4415)$
mesons in collisions with protons. The cross sections are plotted in Figs.
3-12, and are parametrized as
\begin{eqnarray}
\sigma^{\rm unpol}(\sqrt {s})
&=&\frac{\vec{P}^{\prime 2}}{\vec{P}^2}
\left\{a_1 \left( \frac {\sqrt {s} -\sqrt {s_0}} {b_1} \right)^{c_1}
\exp \left[ c_1 \left( 1-\frac {\sqrt {s} -\sqrt {s_0}} {b_1} \right) \right]
\right.
\nonumber \\
&&+ \left.
a_2 \left( \frac {\sqrt {s} -\sqrt {s_0}} {b_2} \right)^{c_2}
\exp \left[ c_2 \left( 1-\frac {\sqrt {s} -\sqrt {s_0}} {b_2} \right) \right]
\right\} ,
\end{eqnarray}
where $\sqrt{s_0}$ is the threshold energy, and $a_1$, $b_1$, $c_1$, $a_2$,
$b_2$, and $c_2$ are parameters. The parameter values are listed in Tables 1-2.
The threshold energy of inelastic $p+\psi (3770)$ ($p+\psi (4040)$,
$p+\psi (4160)$, $p+\psi (4415)$) scattering is the sum of the proton and 
$\psi (3770)$ ($\psi (4040)$, $\psi (4160)$, $\psi (4415)$) masses. At the
threshold energy $\mid \vec{P} \mid$ in Eq. (32) equals zero, but
$\mid \vec{P}^{~\prime} \mid$ does not. The cross section is thus infinite at
the threshold energy. The cross sections in Figs. 3-12 are plotted as 
functions of $\sqrt s$ which equals or is larger than the threshold energy
plus $5 \times 10^{-4}$ GeV.

The reactions considered in the present work are all exothermic. When $\sqrt s$
increases from threshold, the cross sections decrease rapidly, and then change
slowly. In the slowly-changing region the cross sections may be tens of
millibarns. For example, the cross sections for 
$p\psi (3770) \to \Sigma_c^{++} D^{*-}$, 
$p\psi (4040) \to \Lambda_c^+ \bar{D}^{*0}$, 
$p\psi (4160) \to \Sigma_c^{++} D^{*-}$, and
$p\psi (4415) \to \Lambda_c^+ \bar{D}^{*0}$ can reach 20 mb, 30 mb, 21 mb, and
15 mb, respectively.
According to the quantum numbers of $\psi (3770)$, $\psi (4040)$, 
$\psi (4160)$, and $\psi (4415)$ mesons, the numbers of their radial nodes are 
0, 2, 1, and 3, respectively. If there is a node in the radial wave function
$R_{L_B}(r_{c\bar c})$, cancellation between the wave functions on both sides
of the node occurs in the integration involved in the transition amplitudes, 
thus cross sections are reduced. The $\psi (4040)$ mass is near the 
$\psi (4160)$ mass. Since the $\psi (4040)$ meson has one node more than the
$\psi (4160)$ meson, the integration related to $\psi (4040)$ should have more 
cancellation than that related to $\psi (4160)$. However, the wave function
of $\psi (4160)$ contains the spherical harmonics $Y_{2M_B}$ 
($M_B=-2,-1,0,1,2$), and the wave function of $\psi (4040)$ contains the 
constant spherical harmonics $Y_{00}$. Then, the integration related to 
$\psi (4160)$ may have more cancellation than that related to $\psi (4040)$.
Therefore, at the threshold energy plus $5 \times 10^{-4}$ GeV, the cross
sections for $p+\psi(4040)$ reactions are larger in Fig. 3 and Fig. 7
or smaller in Figs. 4-6 and Figs. 8-12 than the ones for
$p+\psi(4160)$ reactions.

The mesonic quark-antiquark relative-motion wave functions are decreasing
functions of the quark-antiquark relative momentum. 
$\phi_{A\rm space}(\vec{p}_\rho, \vec{p}_\lambda)$
($\phi_{C\rm space}(\vec{p}_\rho^{~\prime}, \vec{p}_\lambda^{~\prime})$)
in the transition amplitudes is an
exponentially decreasing function of $\vec{p}_\rho$ and $\vec{p}_\lambda$
($\vec{p}_\rho^{~\prime}$ and $\vec{p}_\lambda^{~\prime}$). The quark-antiquark
relative momenta, $\vec{p}_\rho$, $\vec{p}_\lambda$, $\vec{p}_\rho^{~\prime}$, 
and $\vec{p}_\lambda^{~\prime}$, are given by the expressions enclosed by the
parentheses that follow $\phi_{BJ_BJ_{Bz}}$, $\phi_{D\rm rel}$, 
$\phi_{A\rm space}$, and $\phi_{C\rm space}$ in Eqs. (30) and (31). These
expressions may have $\vec{p}_{q_1q_2q_3,c\bar c}$ and 
$\vec{p}_{q_1q_2c,q_3\bar c}^{~\prime}$. In the center-of-momentum frame of 
the proton and the charmonium, $\vec{p}_{q_1q_2q_3,c\bar c}$ and 
$\vec{p}_{q_1q_2c,q_3\bar c}^{~\prime}$ equal $\vec P$ and $\vec{P}^\prime$,
respectively. Therefore, 
the quark-antiquark relative momenta, $\vec{p}_\rho$, $\vec{p}_\lambda$, 
$\vec{p}_\rho^{~\prime}$, and $\vec{p}_\lambda^{~\prime}$, bear linear relation
to $\vec P$ and $\vec{P}^\prime$. At the threshold energy plus 
$5 \times 10^{-4}$ GeV, $\mid \vec{P} \mid$ almost equals zero, and 
$\mid \vec{P}^\prime \mid$ of any $p+\psi(3770)$ reaction is smaller than 
$\mid \vec{P}^\prime \mid$
of $p+\psi(4040)$, $p+\psi(4160)$, and $p+\psi(4415)$ reactions with the same 
final charmed baryon and the same final charmed meson in any of Figs. 3-12. 
The transition amplitudes for the $p+\psi(3770)$ reaction are larger than those
for the $p+\psi(4040)$, $p+\psi(4160)$, and $p+\psi(4415)$ reactions. 
Therefore, at the threshold energy plus 
$5 \times 10^{-4}$ GeV, the cross section for 
$p\psi(3770) \to \Lambda_c^+ \bar{D}^0$ in Fig. 3 is larger than those for 
$p\psi(4040) \to \Lambda_c^+ \bar{D}^0$, 
$p\psi(4160) \to \Lambda_c^+ \bar{D}^0$, and 
$p\psi(4415) \to \Lambda_c^+ \bar{D}^0$; similar results are displayed in
Figs. 4-12. From the ten figures 
we can also understand that the cross sections for $p+\psi(4160)$
reactions at the threshold energy plus $5 \times 10^{-4}$ GeV are larger
than the ones for $p+\psi(4415)$ reactions.

The measured proton mass has a very small uncertainty, and the uncertainty
is neglected here. The measured masses of charmed baryons have uncertainties
\cite{PDG}, for example, the $\Lambda_c^+$ mass has an error of 0.14 MeV.
The measured mass of every charmed baryon has a maximum and a minimum, for 
example, the $\Lambda_c^+$ mass has the maximum mass 2286.60 MeV and the 
minimum mass
2286.32 MeV. Fits to the maximum experimental masses of $\Lambda_c^+$, 
$\Sigma_c^{++}$, $\Sigma_c^+$, $\Sigma_c^{*++}$, and $\Sigma_c^{*+}$ baryons 
give $\alpha_{\lambda_{\rm f}}=0.43$ GeV and
\begin{displaymath}
\alpha_\rho=0.222571~{\rm GeV},~~~~~~{m_{\Lambda_c^+}=2.28660~{\rm GeV};}
\end{displaymath}
\begin{displaymath}
\alpha_\rho=0.196254~{\rm GeV},~~~~~~{m_{\Sigma_c^{++}}=2.45411~{\rm GeV};}
\end{displaymath}
\begin{displaymath}
\alpha_\rho=0.196365~{\rm GeV},~~~~~~{m_{\Sigma_c^{+}}=2.4533~{\rm GeV};}
\end{displaymath}
\begin{displaymath}
\alpha_\rho=0.190894~{\rm GeV},~~~~~~{m_{\Sigma_c^{*++}}=2.51862~{\rm GeV};}
\end{displaymath}
\begin{displaymath}
\alpha_\rho=0.19076~{\rm GeV},~~~~~~{m_{\Sigma_c^{*+}}=2.5198~{\rm GeV}.}
\end{displaymath}
Using these values of $\alpha_\rho$ and $\alpha_{\lambda_{\rm f}}$, we obtain
unpolarized cross sections which are denoted as $\sigma_{\rm lm}^{\rm unpol}$.
The differences between $\sigma_{\rm lm}^{\rm unpol}$ and $\sigma^{\rm unpol}$
shown in Figs. 3-12 are plotted as
the lower solid, dashed, dotted, and dot-dashed curves in Figs. 13-22.
Fits to the minimum experimental masses of $\Lambda_c^+$, 
$\Sigma_c^{++}$, $\Sigma_c^+$, $\Sigma_c^{*++}$, and $\Sigma_c^{*+}$ baryons 
give $\alpha_{\lambda_{\rm f}}=0.43$ GeV and
\begin{displaymath}
\alpha_\rho=0.222617~{\rm GeV},~~~~~~{m_{\Lambda_c^+}=2.28632~{\rm GeV};}
\end{displaymath}
\begin{displaymath}
\alpha_\rho=0.196292~{\rm GeV},~~~~~~{m_{\Sigma_c^{++}}=2.45383~{\rm GeV};}
\end{displaymath}
\begin{displaymath}
\alpha_\rho=0.196475~{\rm GeV},~~~~~~{m_{\Sigma_c^{+}}=2.4525~{\rm GeV};}
\end{displaymath}
\begin{displaymath}
\alpha_\rho=0.190946~{\rm GeV},~~~~~~{m_{\Sigma_c^{*++}}=2.51822~{\rm GeV};}
\end{displaymath}
\begin{displaymath}
\alpha_\rho=0.19136~{\rm GeV},~~~~~~{m_{\Sigma_c^{*+}}=2.5152~{\rm GeV}.}
\end{displaymath}
Using these values of $\alpha_\rho$ and $\alpha_{\lambda_{\rm f}}$, we obtain
unpolarized cross sections which are denoted as $\sigma_{\rm sm}^{\rm unpol}$.
The differences between $\sigma_{\rm sm}^{\rm unpol}$ and 
$\sigma^{\rm unpol}$ are plotted as
the upper solid, dashed, dotted, and dot-dashed  curves in Figs. 13-22.
In every figure the orange (green, red, blue) band between the lower and upper
solid (dashed, dotted, dot-dashed) curves show uncertainties of the unpolarized
cross sections, which are labeled as $\sigma_{\rm uncer}$ and are caused by
the uncertainties of the $\alpha_\rho$ value. However,
the cross-section uncertainties are too small to be shown if the bands are 
attached to those curves in Figs. 3-12. The cross-section 
uncertainties are small because of the small uncertainties of the $\alpha_\rho$
values, which correspond to errors of measurement of the baryon masses.

In Refs. \cite{GI,BGS,VFV} $\psi (3770)$, $\psi (4040)$, $\psi (4160)$, and 
$\psi (4415)$ mesons are identified with the $1 ^3D_1$, $3 ^3S_1$, $2 ^3D_1$, 
and $4 ^3S_1$ states of a charm quark and a charm antiquark. This 
identification is also true with the potential given in Eq. (33), and we then
study inelastic scattering of a nucleon by the four $c\bar c$ mesons in the 
present work. However, we note that the quantum states of $\psi (3770)$ and 
$\psi (4415)$ mesons are open to debate. The $1 ^3D_1$ $c\bar c$ state of the 
$\psi (3770)$ meson is suggested to be
mixed with the $2 ^3S_1$ $c\bar c$ state in Ref. \cite{Rosner}, and may contain
a four-quark component with the up- and down-quarks and antiquarks in Ref. 
\cite{Voloshin}. The $\psi (4415)$ meson may be a $5 ^3S_1$ $c\bar c$ state 
given in the screened potential model \cite{LC}, a $3 ^3D_1$ $c\bar c$ state 
obtained with a quark potential derivd from a Lagrangian with chiral symmetry 
breaking in Ref. \cite{Segovia}, or a $c\bar c$ hybrid recognized in lattice 
calculations of meson masses \cite{lathybrid}, from the 
nonrelativistic reduction of the QCD Hamiltonian in the Coulomb gauge 
\cite{GSGVS}, and in the flux-tube model \cite{IKP}.

\vspace{0.5cm}
\leftline{\bf IV. SUMMARY }
\vspace{0.5cm}

Flavor interchange between a nucleon and a $c\bar c$ meson breaks the meson.
According to the quark interchange mechanism, we have derived formulas of
the transition amplitudes that include wave functions and 
constituent-constituent potentials. The 
transition amplitudes are used to calculate unpolarized cross sections for the
reactions: 
$pR \to \Lambda_c^+ \bar{D}^0$, $pR \to \Lambda_c^+ \bar{D}^{*0}$,
$pR \to \Sigma_c^{++} D^-$, $pR \to \Sigma_c^{++} D^{*-}$,
$pR \to \Sigma_c^{+} \bar{D}^0$, $pR \to \Sigma_c^{+} \bar{D}^{*0}$,
$pR \to \Sigma_c^{*++} D^-$, $pR \to \Sigma_c^{*++} D^{*-}$,
$pR \to \Sigma_c^{*+} \bar{D}^0$, and $pR \to \Sigma_c^{*+} \bar{D}^{*0}$,
where $R$ represents $\psi (3770)$, $\psi (4040)$, $\psi (4160)$, or 
$\psi (4415)$. These reactions are exothermic, and the $\sqrt s$ dependence of
their cross sections is so that the cross sections decrease rapidly near
threshold and change slowly when the center-of-mass energy of the nucleon and 
the $c\bar c$ meson is not close to threshold. In the slowly-changing region
the cross sections may be tens of millibarns. The cross sections also depend
on nodes in the radial wave functions of the $c\bar c$ mesons. Numerical cross
sections are parametrized. Cross sections
for reactions of a neutron and a $c\bar c$ meson are obtained from those
of a proton and the $c\bar c$ meson.

\vspace{0.5cm}
\leftline{\bf ACKNOWLEDGEMENTS}
\vspace{0.5cm}

This work was supported by the project STRONG-2020 of European Center for
Theoretical Studies in Nuclear Physics and Related Areas.

\newpage
\begin{figure}[htbp]
  \centering
    \includegraphics[scale=0.65]{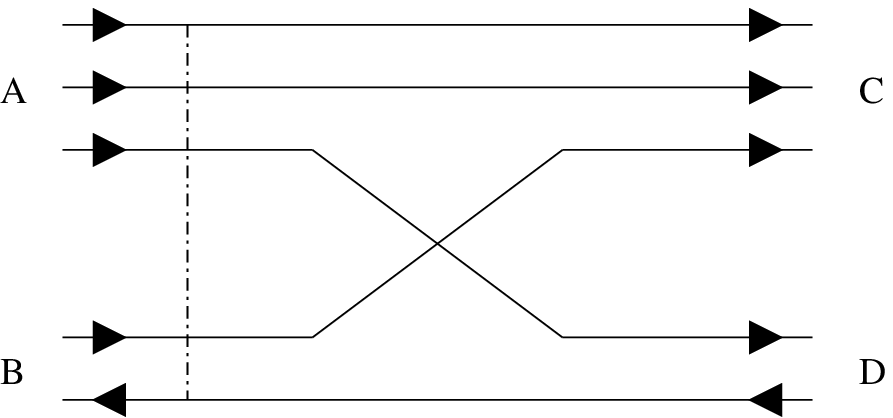}
      \hspace{2cm}
    \includegraphics[scale=0.65]{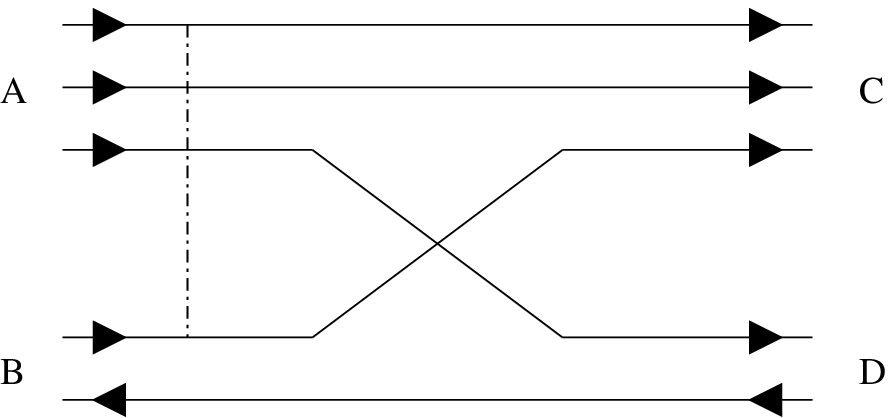}
      \vskip 72pt
    \includegraphics[scale=0.65]{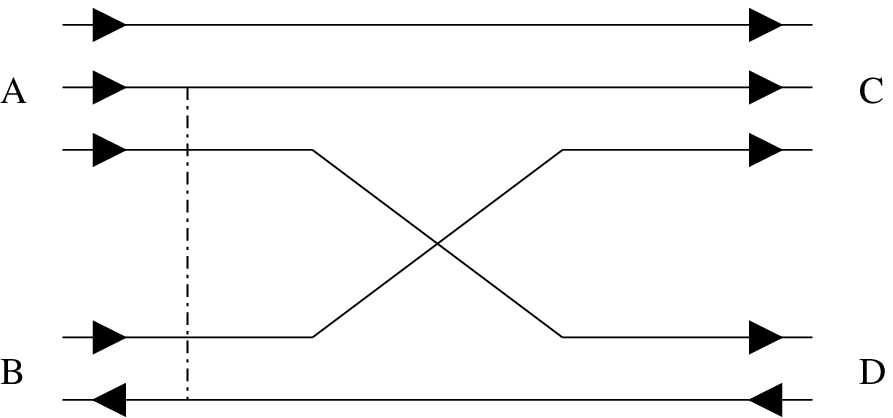}
      \hspace{2cm}
    \includegraphics[scale=0.65]{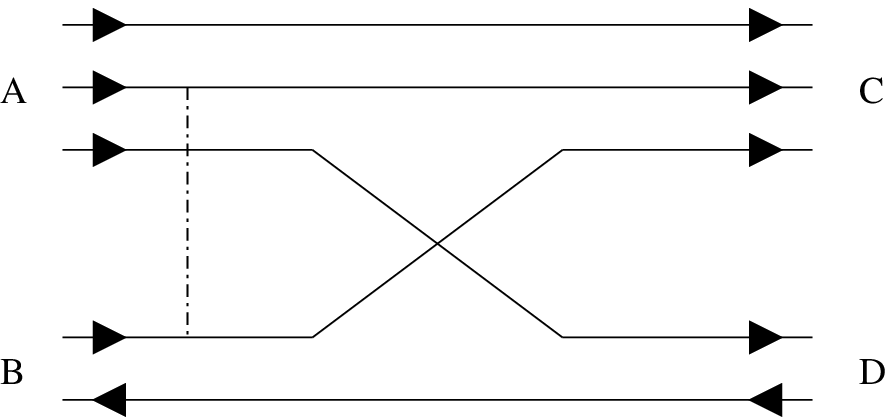}
      \vskip 72pt
    \includegraphics[scale=0.65]{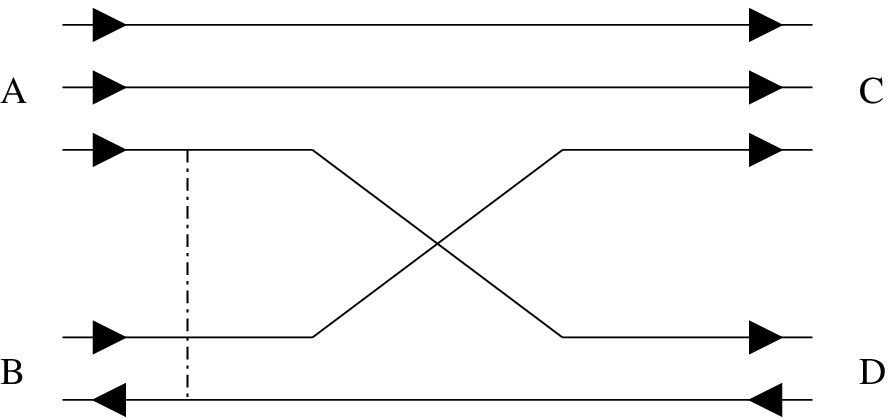}
      \hspace{2cm}
    \includegraphics[scale=0.65]{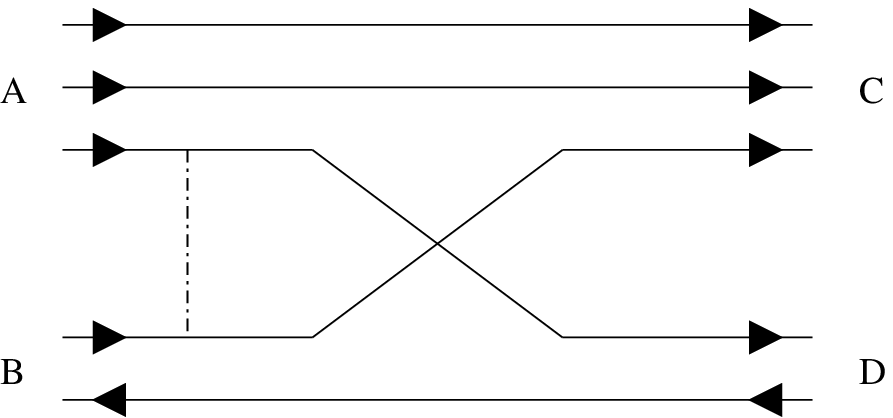}
\caption{Scattering in the prior form. Solid lines with triangles right (left) 
represent quarks (antiquarks). Dot-dashed lines indicate interactions.}
\label{fig1}
\end{figure}

\newpage
\begin{figure}[htbp]
  \centering
    \includegraphics[scale=0.65]{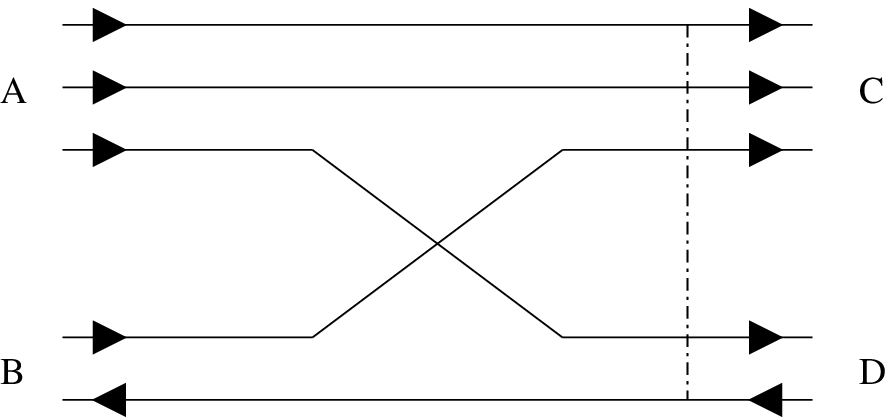}
      \hspace{2cm}
    \includegraphics[scale=0.65]{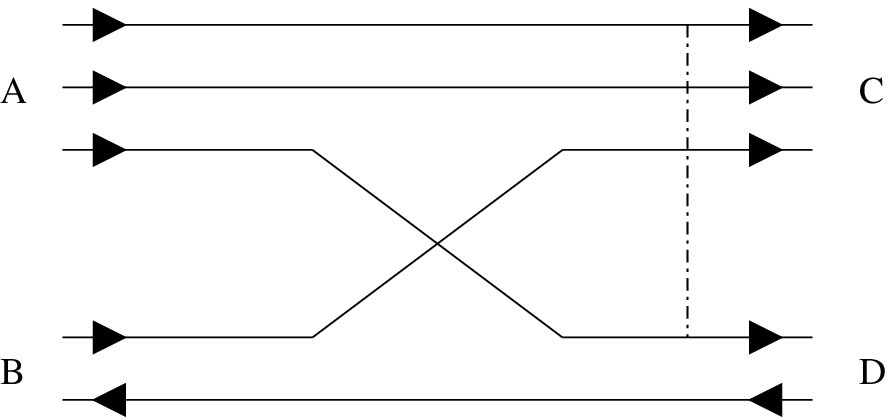}
      \vskip 72pt
    \includegraphics[scale=0.65]{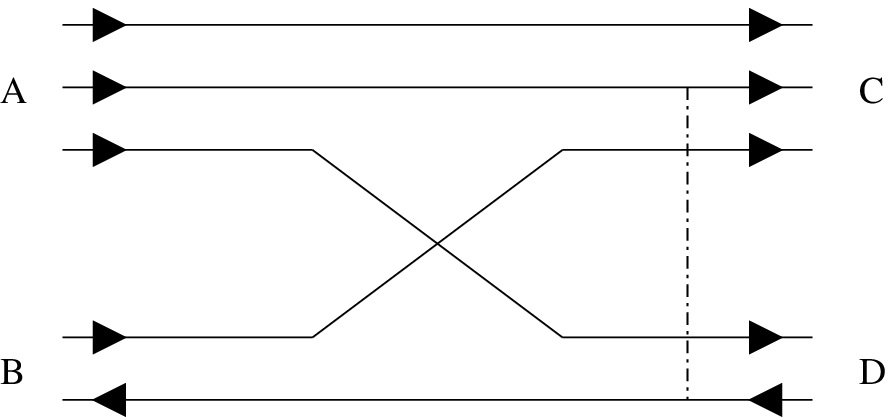}
      \hspace{2cm}
    \includegraphics[scale=0.65]{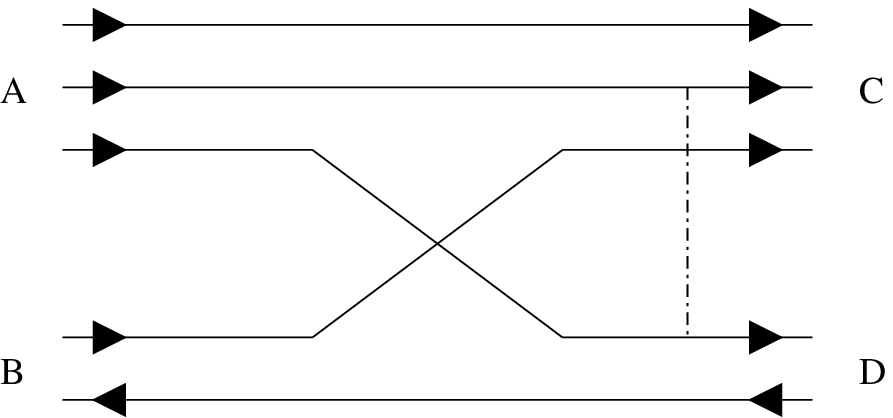}
      \vskip 72pt
    \includegraphics[scale=0.65]{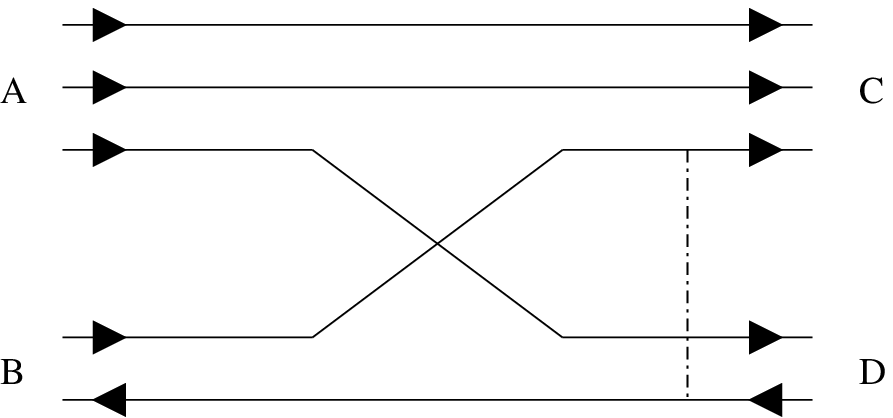}
      \hspace{2cm}
    \includegraphics[scale=0.65]{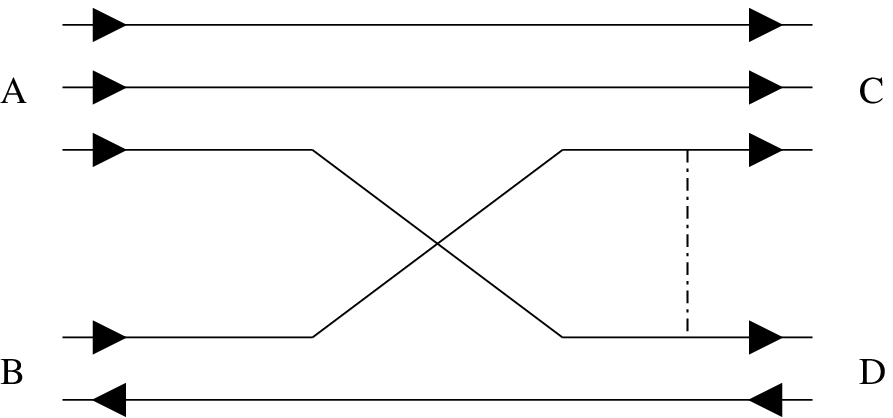}
\caption{Scattering in the post form. Solid lines with triangles right (left) 
represent quarks (antiquarks). Dot-dashed lines indicate interactions.}
\label{fig2}
\end{figure}

\newpage
\begin{figure}[htbp]
  \centering
    \includegraphics[scale=0.65]{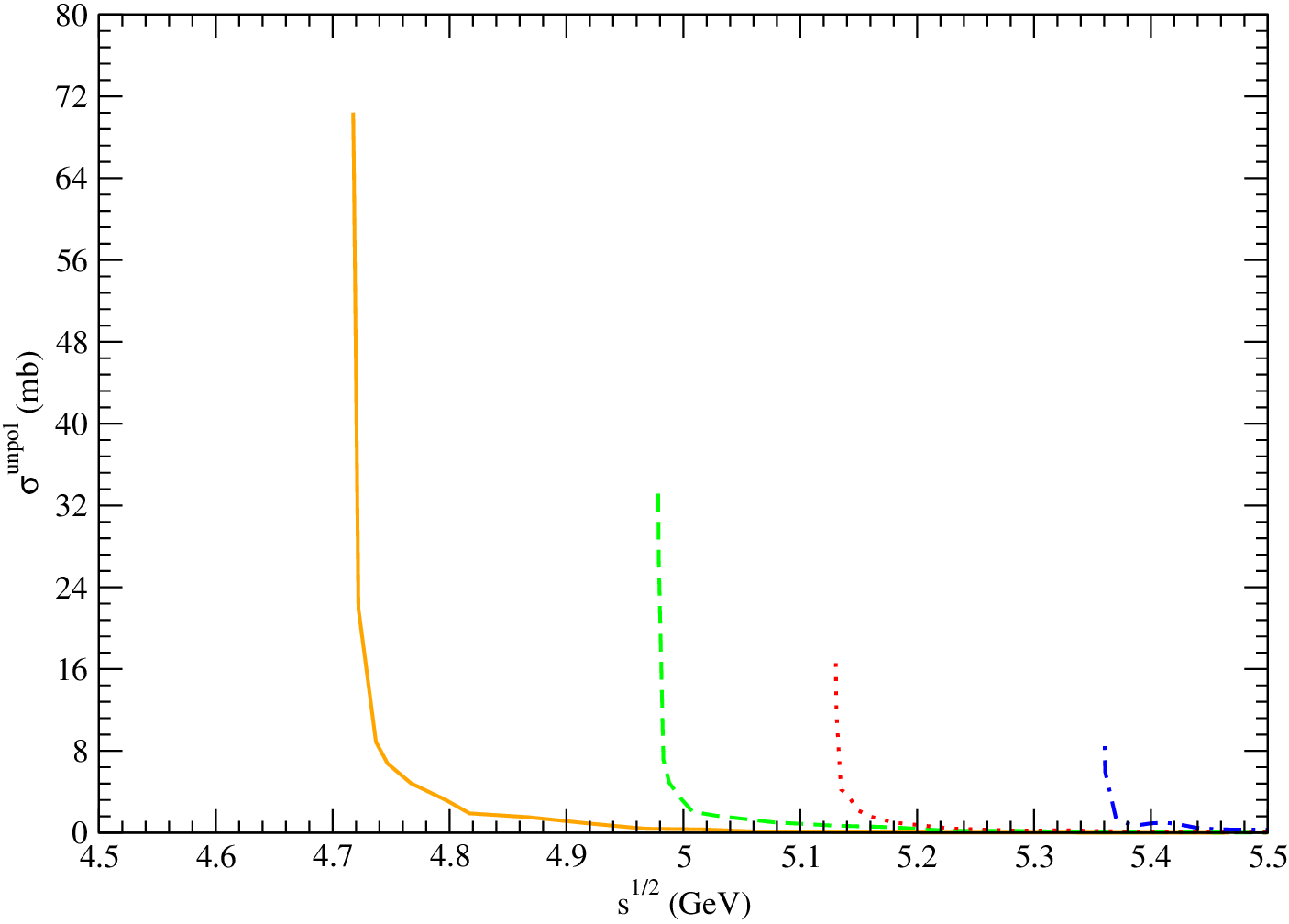}
\caption{Solid, dashed, dotted, and dot-dashed curves stand for cross sections
for $p \psi(3770) \to \Lambda_c^+ \bar{D}^0$, 
$p \psi(4040) \to \Lambda_c^+ \bar{D}^0$, 
$p \psi(4160) \to \Lambda_c^+ \bar{D}^0$, and 
$p \psi(4415) \to \Lambda_c^+ \bar{D}^0$, respectively.}
\label{fig3}
\end{figure}

\newpage
\begin{figure}[htbp]
  \centering
    \includegraphics[scale=0.65]{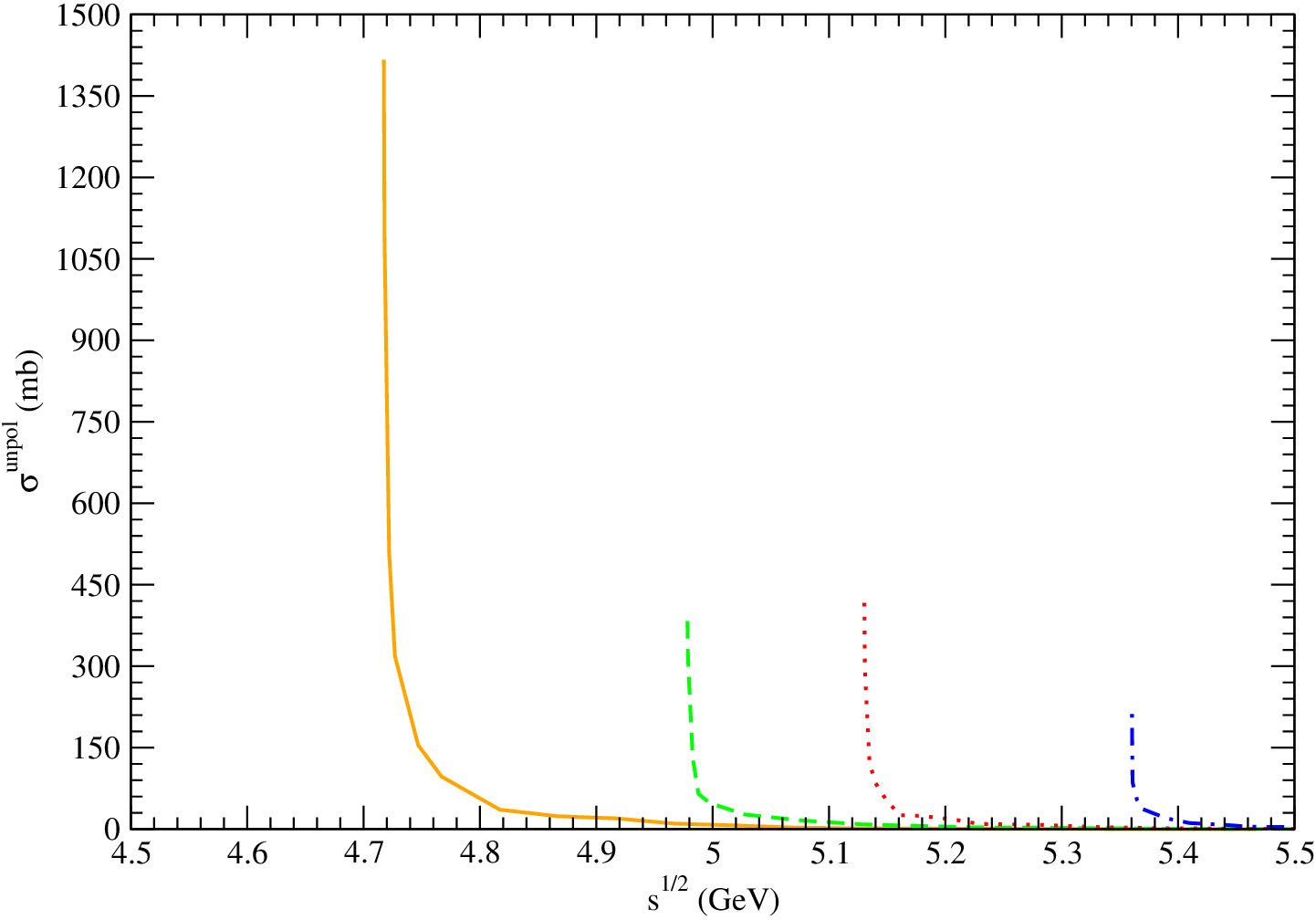}
\caption{Solid, dashed, dotted, and dot-dashed curves stand for cross sections
for $p \psi(3770) \to \Lambda_c^+ \bar{D}^{*0}$, 
$p \psi(4040) \to \Lambda_c^+ \bar{D}^{*0}$, 
$p \psi(4160) \to \Lambda_c^+ \bar{D}^{*0}$, and 
$p \psi(4415) \to \Lambda_c^+ \bar{D}^{*0}$, respectively.}
\label{fig4}
\end{figure}

\newpage
\begin{figure}[htbp]
  \centering
    \includegraphics[scale=0.65]{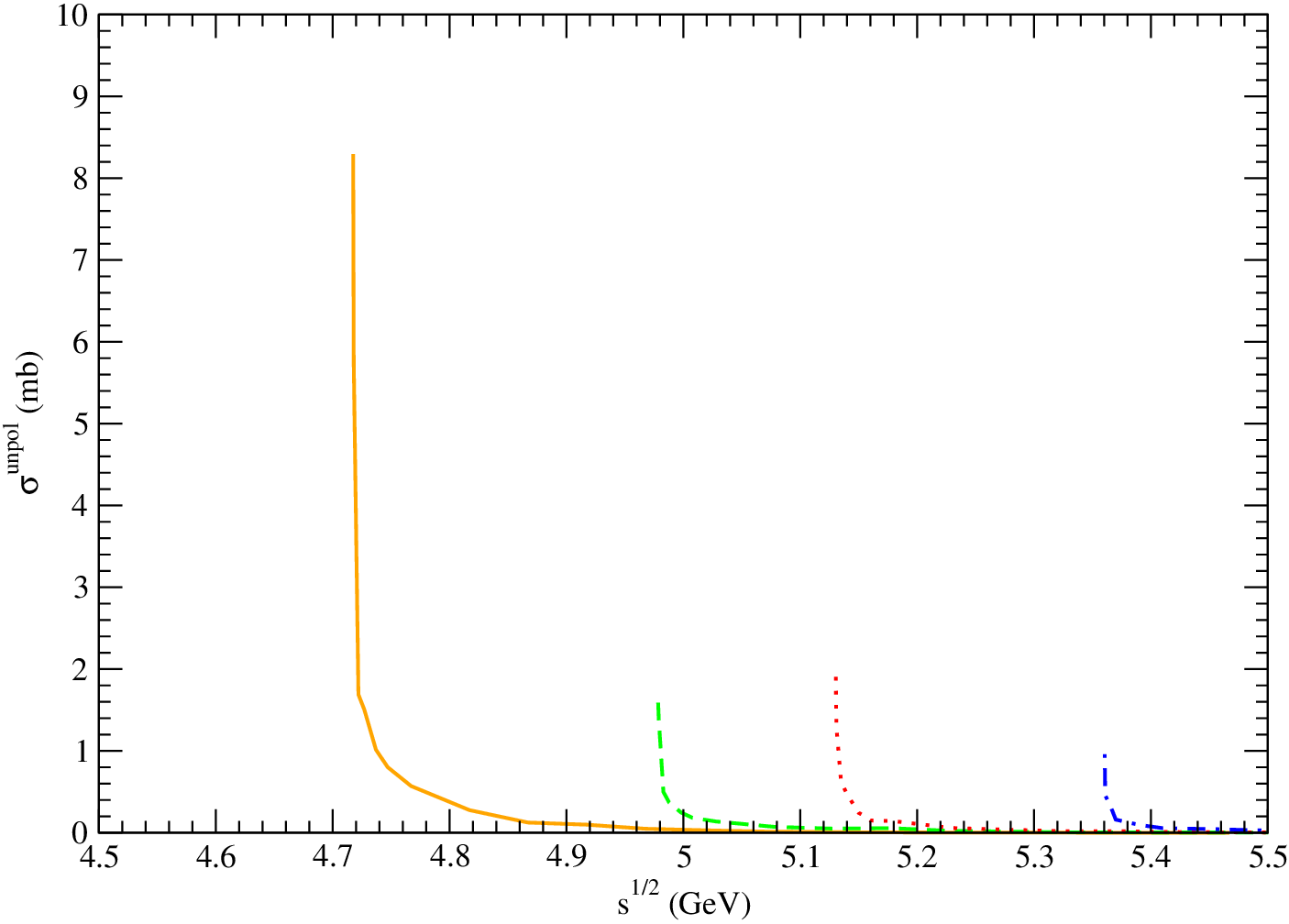}
\caption{Solid, dashed, dotted, and dot-dashed curves stand for cross sections
for $p \psi(3770) \to \Sigma_c^{++} D^-$, 
$p \psi(4040) \to \Sigma_c^{++} D^-$, 
$p \psi(4160) \to \Sigma_c^{++} D^-$, and 
$p \psi(4415) \to \Sigma_c^{++} D^-$, respectively.}
\label{fig5}
\end{figure}

\newpage
\begin{figure}[htbp]
  \centering
    \includegraphics[scale=0.65]{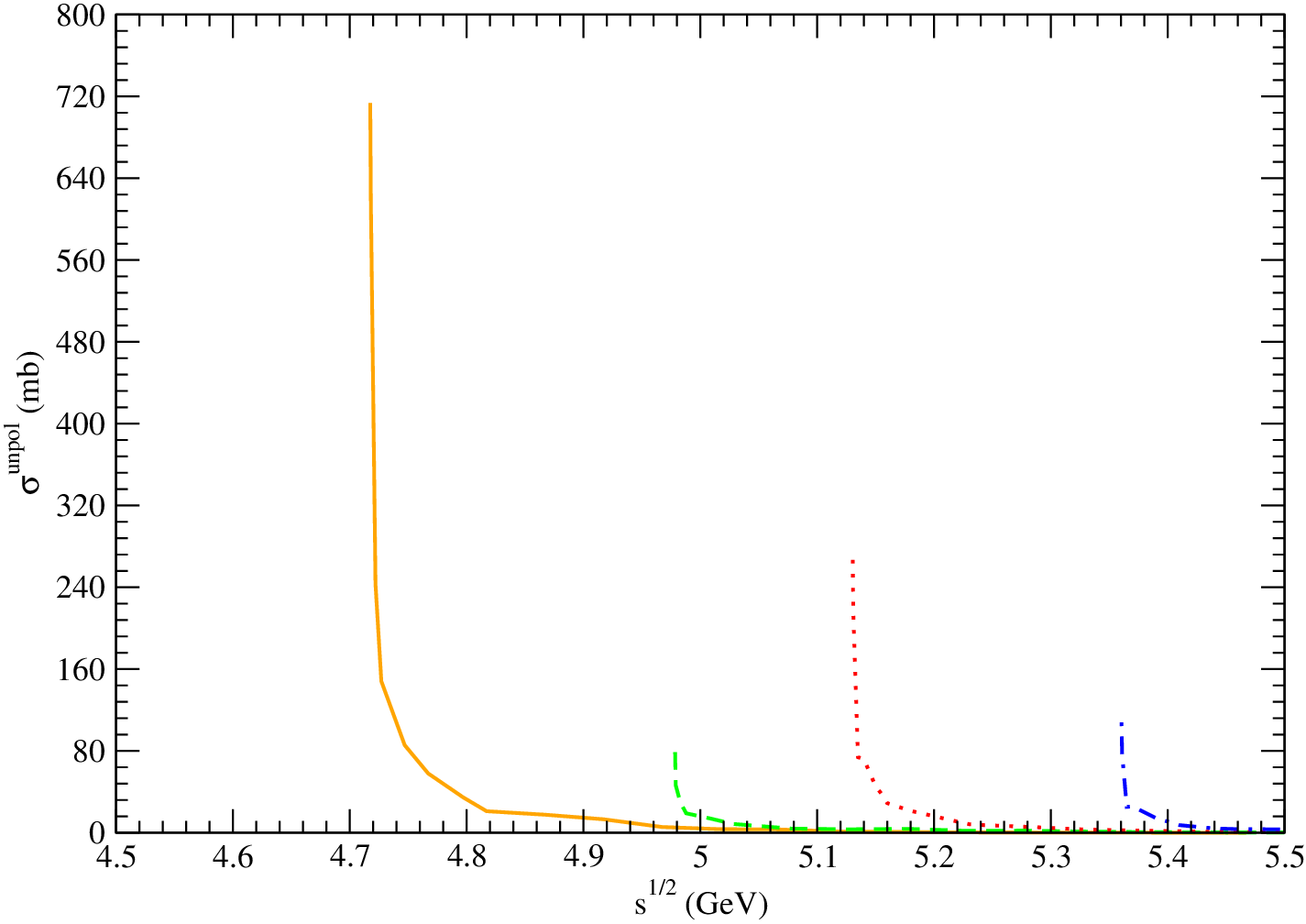}
\caption{Solid, dashed, dotted, and dot-dashed curves stand for cross sections
for $p \psi(3770) \to \Sigma_c^{++} D^{*-}$, 
$p \psi(4040) \to \Sigma_c^{++} D^{*-}$, 
$p \psi(4160) \to \Sigma_c^{++} D^{*-}$, and 
$p \psi(4415) \to \Sigma_c^{++} D^{*-}$, respectively.}
\label{fig6}
\end{figure}

\newpage
\begin{figure}[htbp]
  \centering
    \includegraphics[scale=0.65]{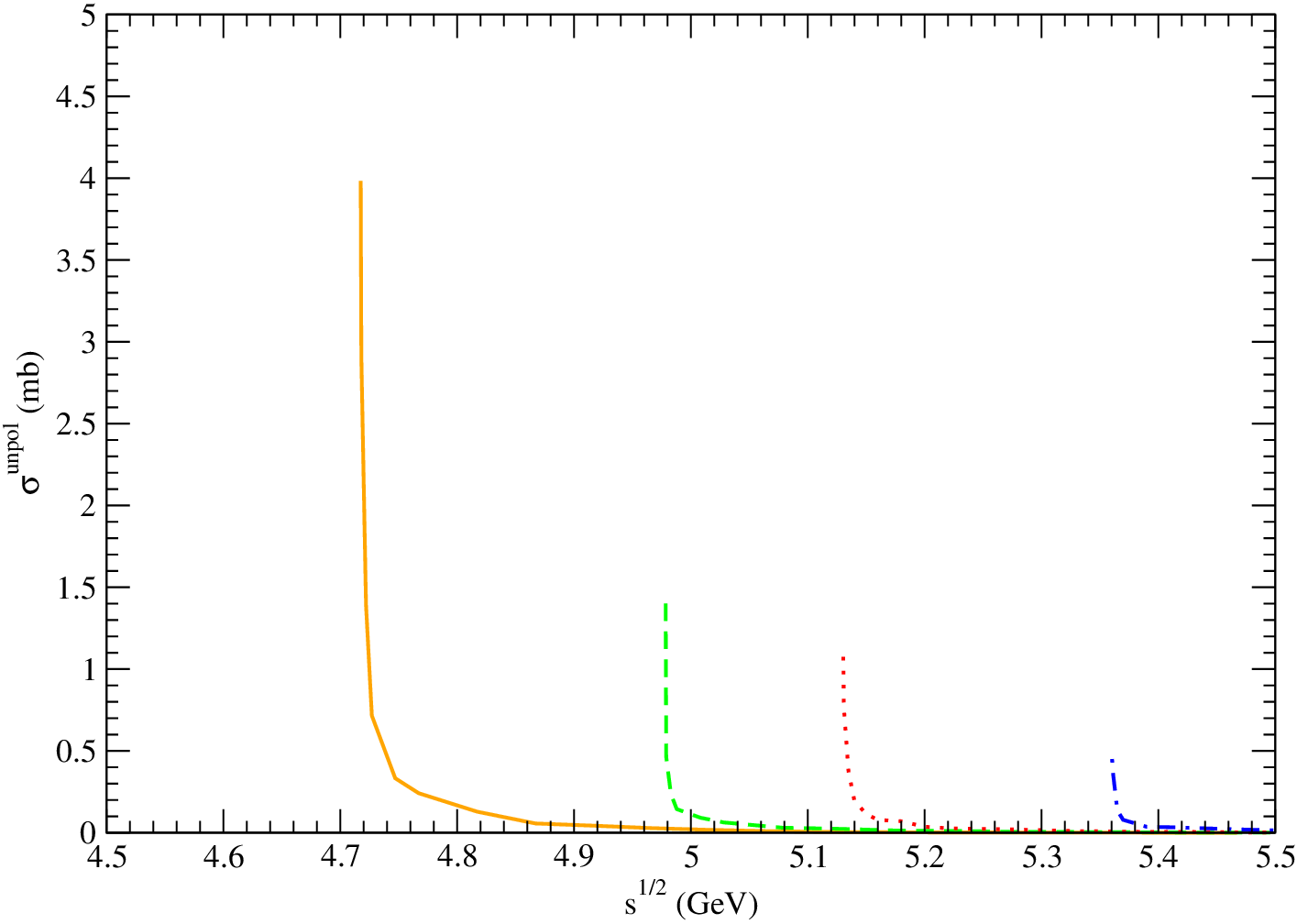}
\caption{Solid, dashed, dotted, and dot-dashed curves stand for cross sections
for $p \psi(3770) \to \Sigma_c^{+} \bar{D}^0$, 
$p \psi(4040) \to \Sigma_c^{+} \bar{D}^0$, 
$p \psi(4160) \to \Sigma_c^{+} \bar{D}^0$, and 
$p \psi(4415) \to \Sigma_c^{+} \bar{D}^0$, respectively.}
\label{fig7}
\end{figure}

\newpage
\begin{figure}[htbp]
  \centering
    \includegraphics[scale=0.65]{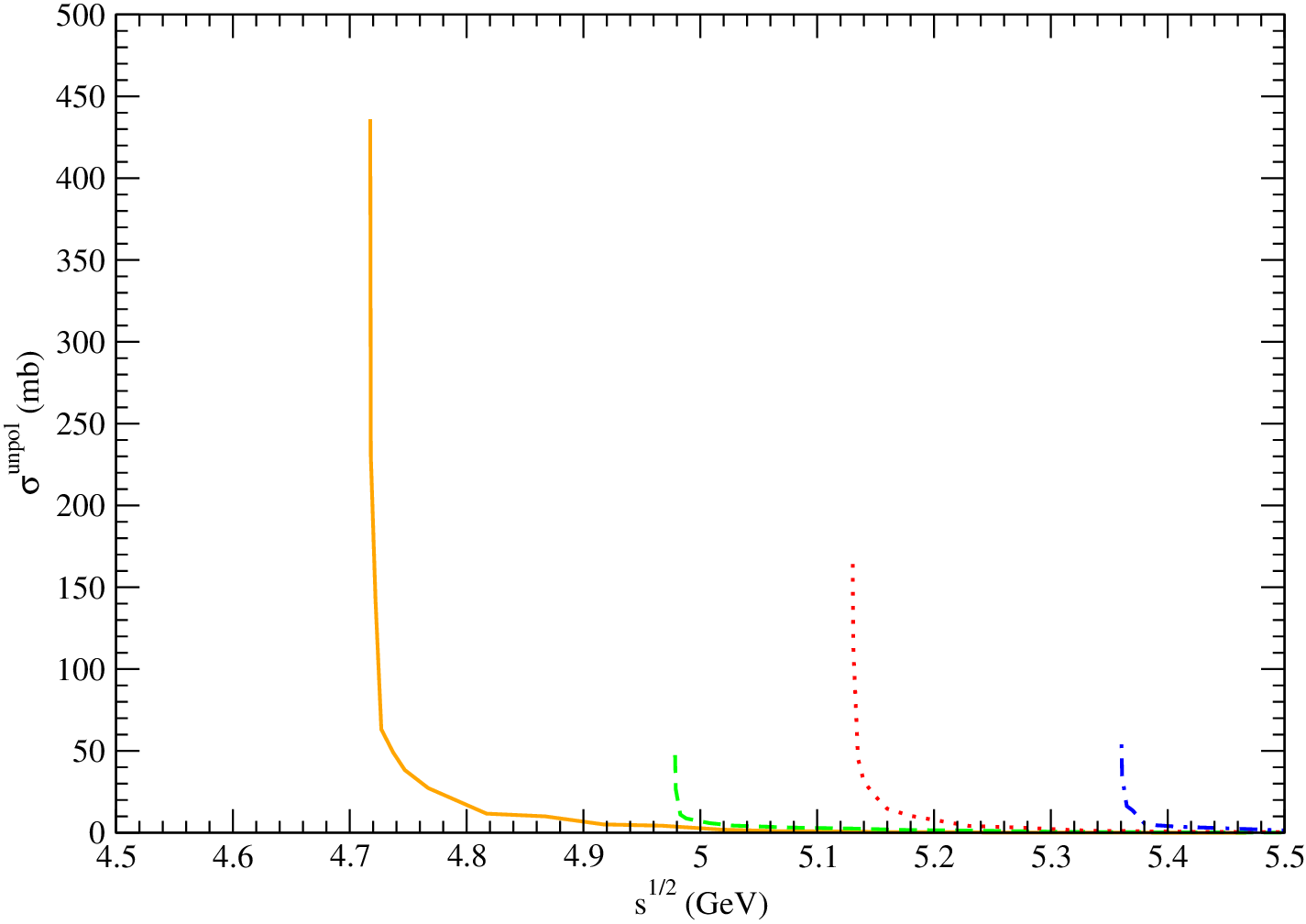}
\caption{Solid, dashed, dotted, and dot-dashed curves stand for cross sections
for $p \psi(3770) \to \Sigma_c^{+} \bar{D}^{*0}$, 
$p \psi(4040) \to \Sigma_c^{+} \bar{D}^{*0}$, 
$p \psi(4160) \to \Sigma_c^{+} \bar{D}^{*0}$, and 
$p \psi(4415) \to \Sigma_c^{+} \bar{D}^{*0}$, respectively.}
\label{fig8}
\end{figure}

\newpage
\begin{figure}[htbp]
  \centering
    \includegraphics[scale=0.65]{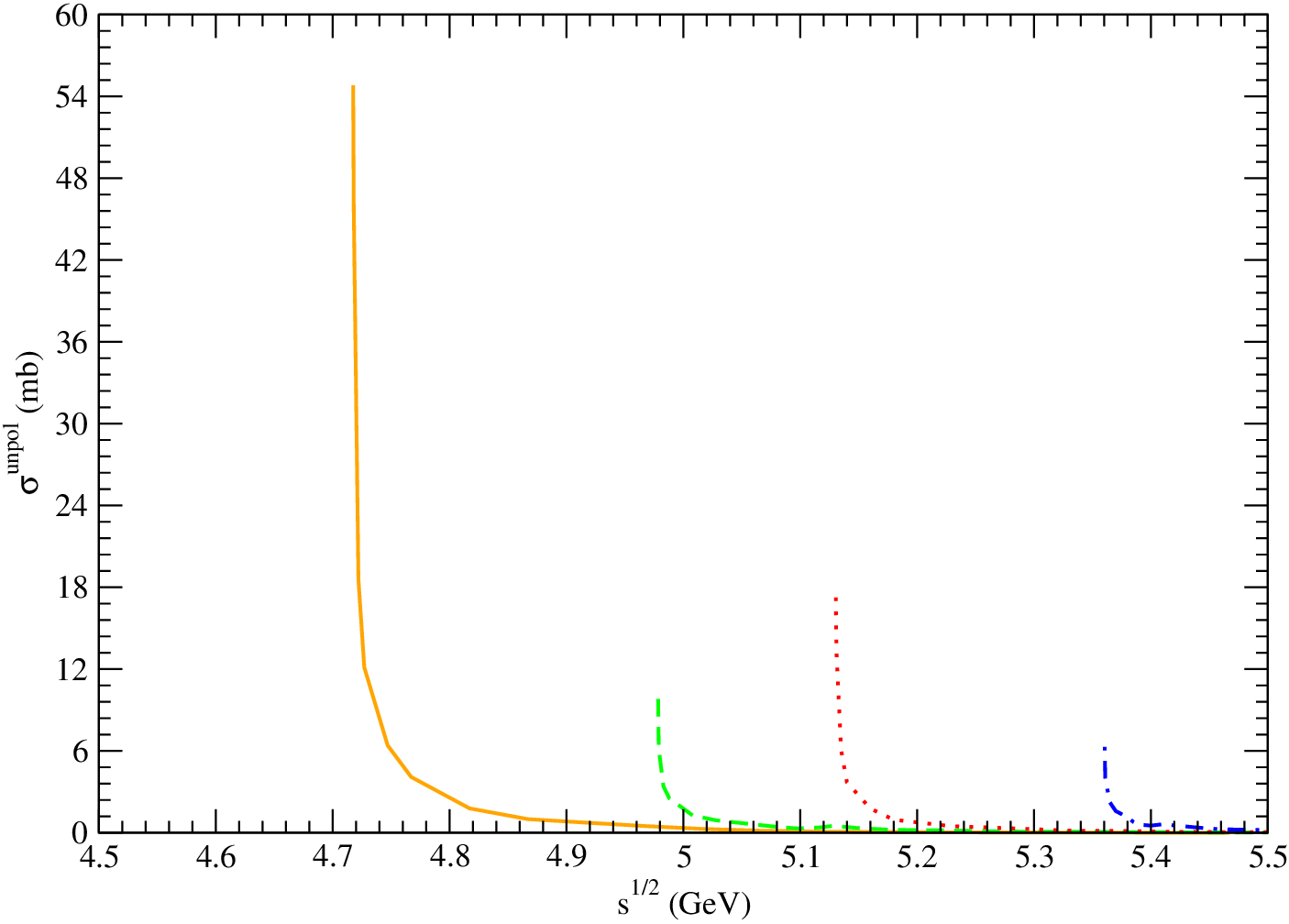}
\caption{Solid, dashed, dotted, and dot-dashed curves stand for cross sections
for $p \psi(3770) \to \Sigma_c^{*++} D^-$, 
$p \psi(4040) \to \Sigma_c^{*++} D^-$, 
$p \psi(4160) \to \Sigma_c^{*++} D^-$, and 
$p \psi(4415) \to \Sigma_c^{*++} D^-$, respectively.}
\label{fig9}
\end{figure}

\newpage
\begin{figure}[htbp]
  \centering
    \includegraphics[scale=0.65]{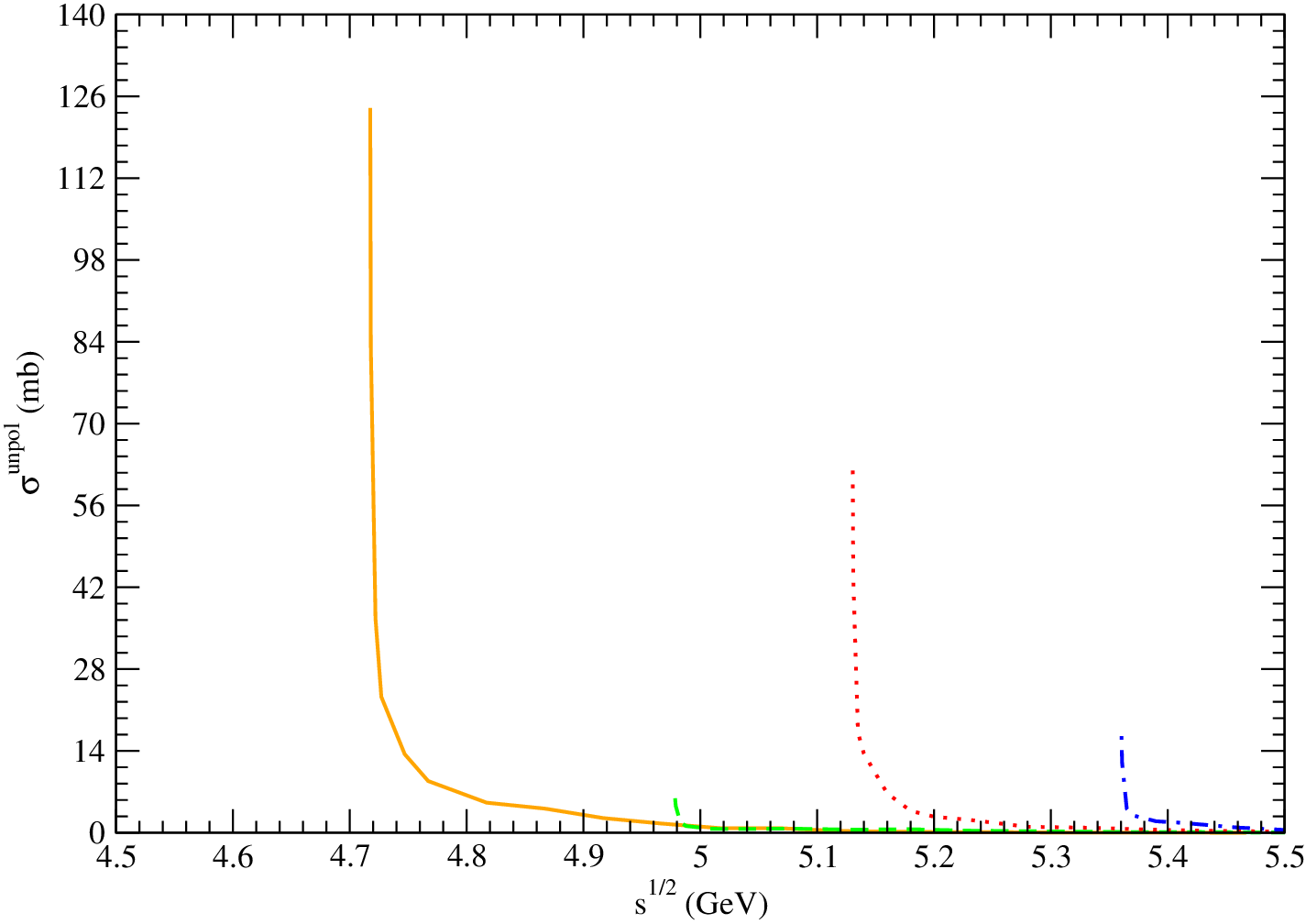}
\caption{Solid, dashed, dotted, and dot-dashed curves stand for cross sections
for $p \psi(3770) \to \Sigma_c^{*++} D^{*-}$, 
$p \psi(4040) \to \Sigma_c^{*++} D^{*-}$, 
$p \psi(4160) \to \Sigma_c^{*++} D^{*-}$, and 
$p \psi(4415) \to \Sigma_c^{*++} D^{*-}$, respectively.}
\label{fig10}
\end{figure}

\newpage
\begin{figure}[htbp]
  \centering
    \includegraphics[scale=0.65]{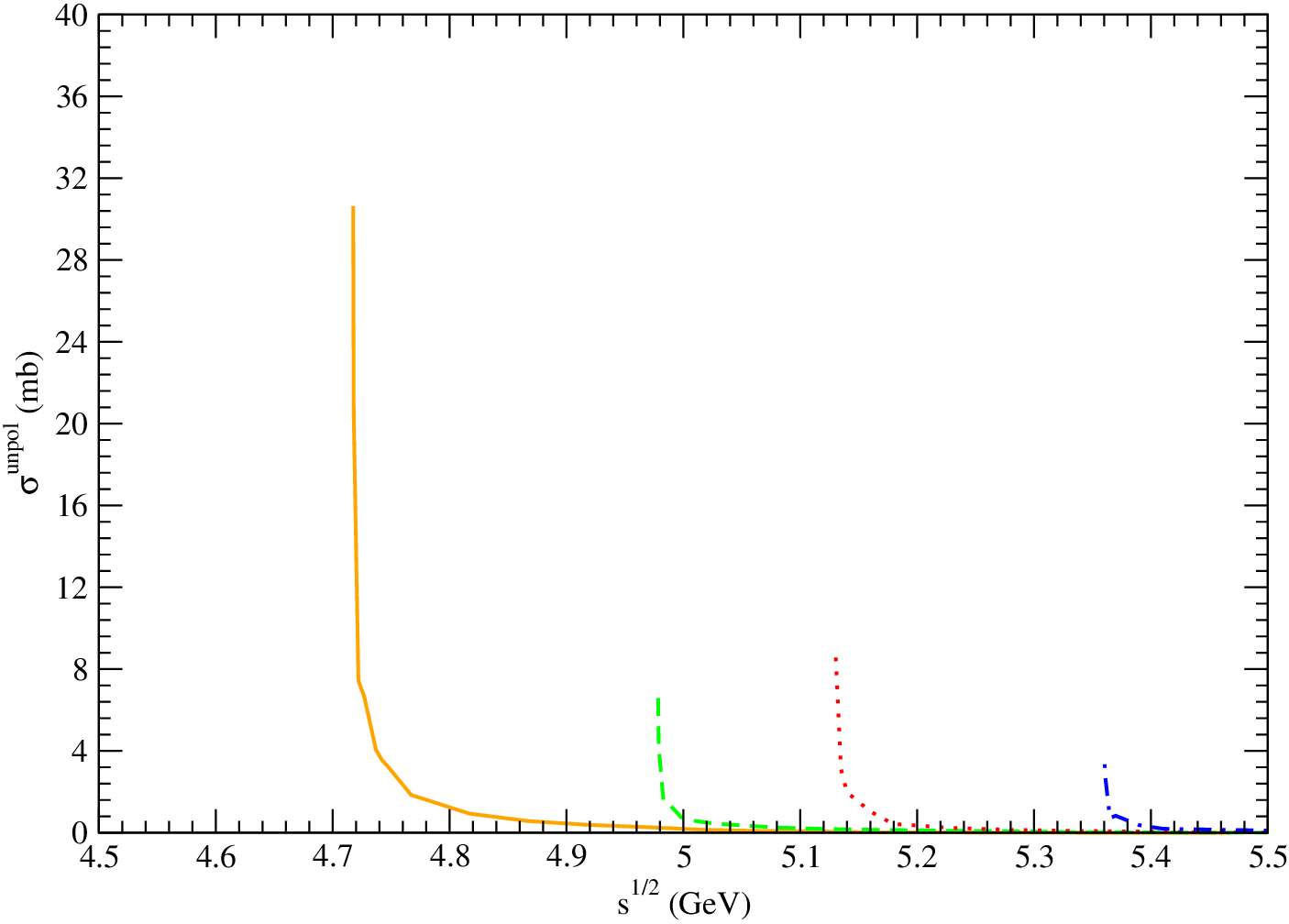}
\caption{Solid, dashed, dotted, and dot-dashed curves stand for cross sections
for $p \psi(3770) \to \Sigma_c^{*+} \bar{D}^0$, 
$p \psi(4040) \to \Sigma_c^{*+} \bar{D}^0$, 
$p \psi(4160) \to \Sigma_c^{*+} \bar{D}^0$, and 
$p \psi(4415) \to \Sigma_c^{*+} \bar{D}^0$, respectively.}
\label{fig11}
\end{figure}

\newpage
\begin{figure}[htbp]
  \centering
    \includegraphics[scale=0.65]{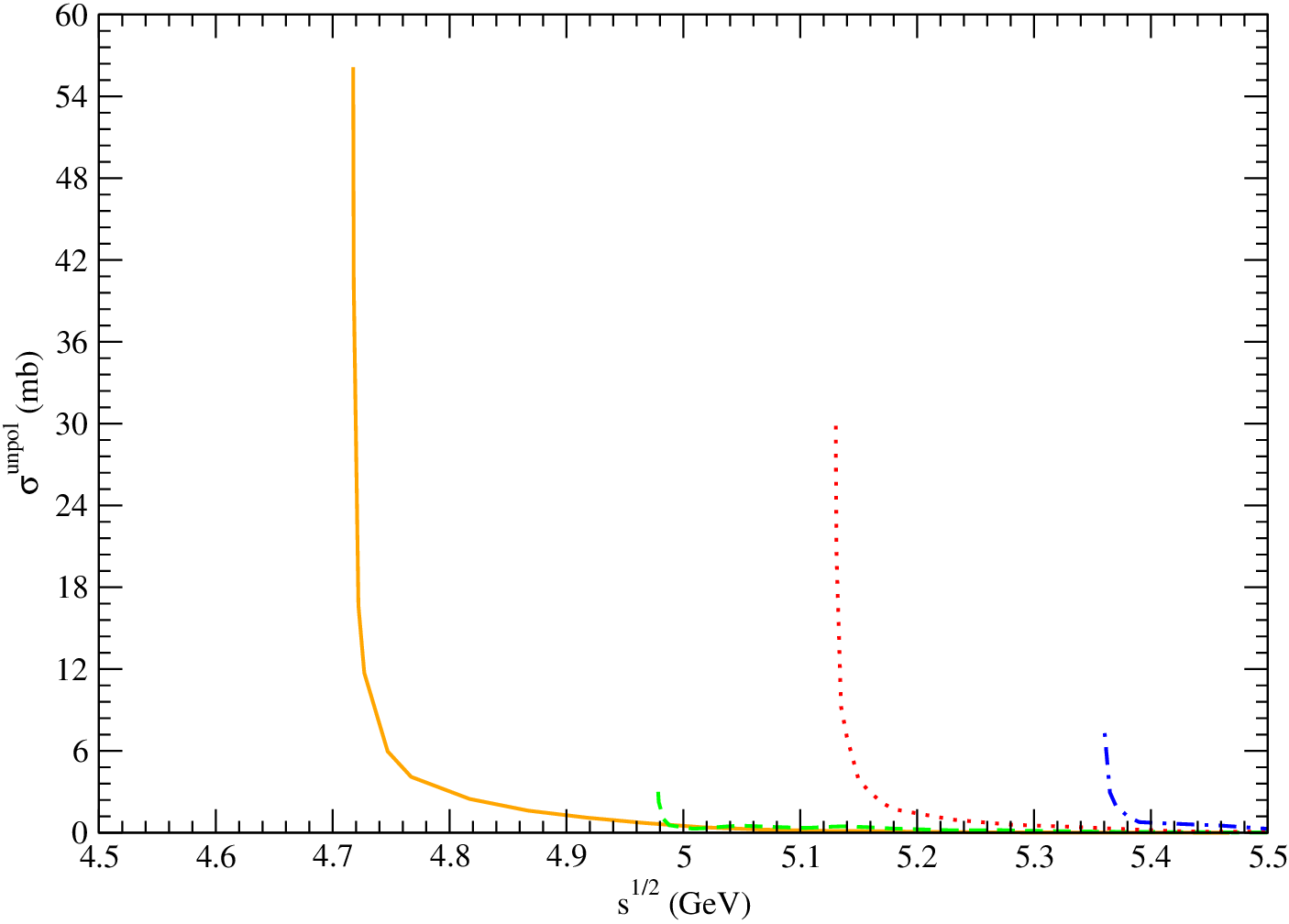}
\caption{Solid, dashed, dotted, and dot-dashed curves stand for cross sections
for $p \psi(3770) \to \Sigma_c^{*+} \bar{D}^{*0}$, 
$p \psi(4040) \to \Sigma_c^{*+} \bar{D}^{*0}$, 
$p \psi(4160) \to \Sigma_c^{*+} \bar{D}^{*0}$, and 
$p \psi(4415) \to \Sigma_c^{*+} \bar{D}^{*0}$, respectively.}
\label{fig12}
\end{figure}

\newpage
\begin{figure}[htbp]
  \centering
    \includegraphics[scale=0.64]{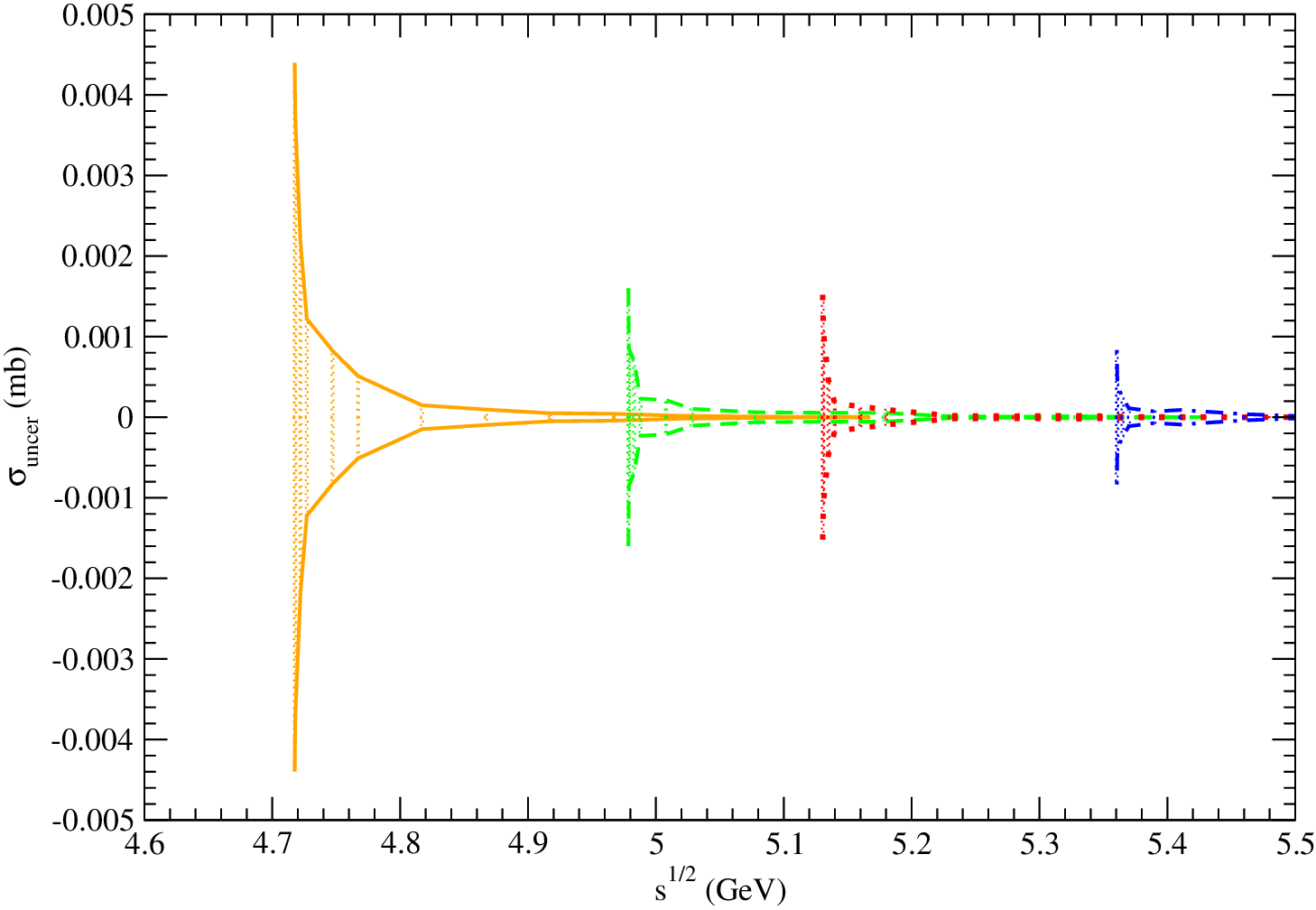}
\caption{The error band between the two solid (dashed, dotted, and dot-dashed)
curves indicates uncertainties of the unpolarized cross sections
for $p \psi(3770) \to \Lambda_c^+ \bar{D}^0$
($p \psi(4040) \to \Lambda_c^+ \bar{D}^0$, 
$p \psi(4160) \to \Lambda_c^+ \bar{D}^0$, and 
$p \psi(4415) \to \Lambda_c^+ \bar{D}^0$).}
\label{fig13}
\end{figure}

\newpage
\begin{figure}[htbp]
  \centering
    \includegraphics[scale=0.65]{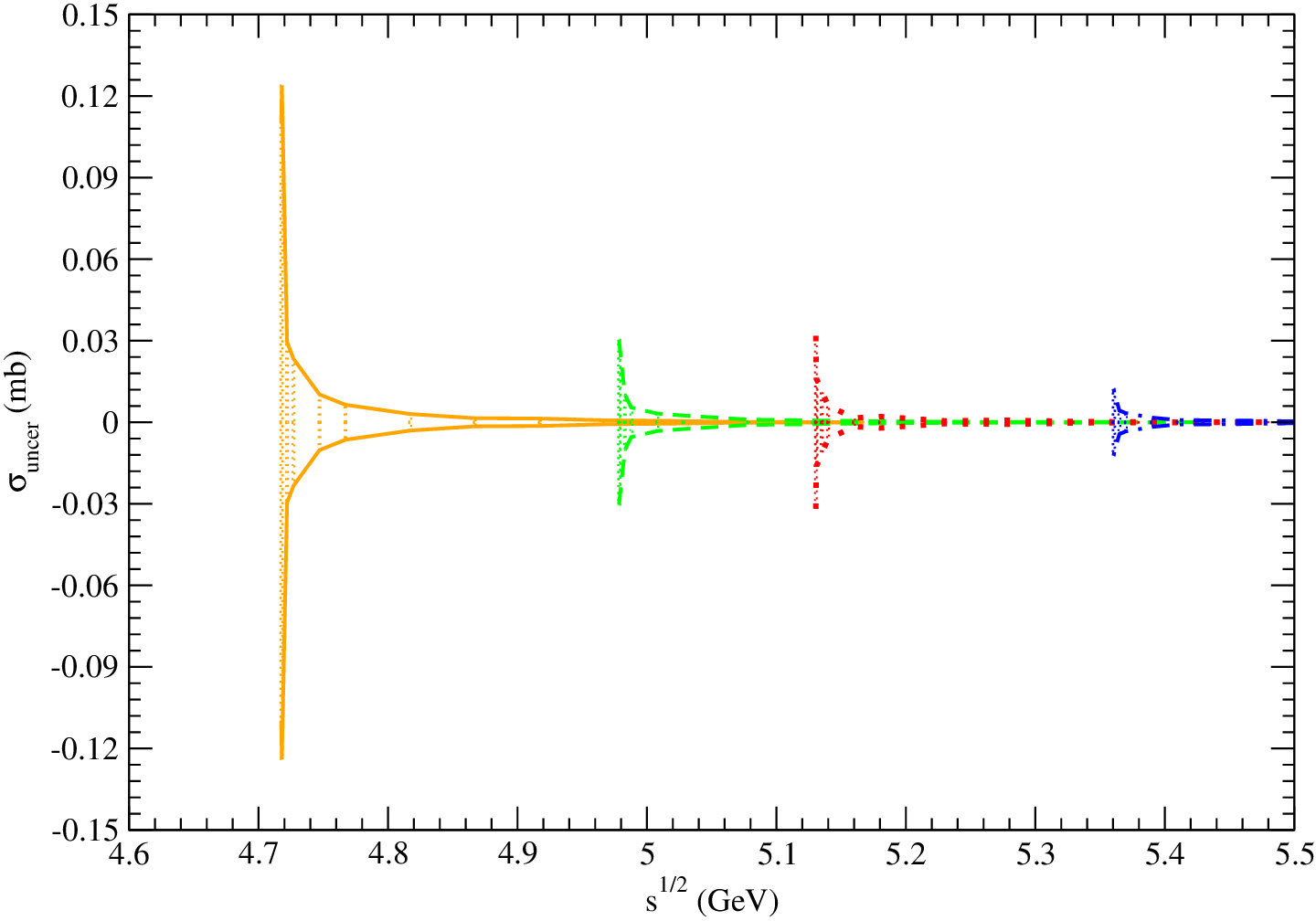}
\caption{The error band between the two solid (dashed, dotted, and dot-dashed)
curves indicates uncertainties of the unpolarized cross sections
for $p \psi(3770) \to \Lambda_c^+ \bar{D}^{*0}$
($p \psi(4040) \to \Lambda_c^+ \bar{D}^{*0}$, 
$p \psi(4160) \to \Lambda_c^+ \bar{D}^{*0}$, and 
$p \psi(4415) \to \Lambda_c^+ \bar{D}^{*0}$).}
\label{fig14}
\end{figure}

\newpage
\begin{figure}[htbp]
  \centering
    \includegraphics[scale=0.63]{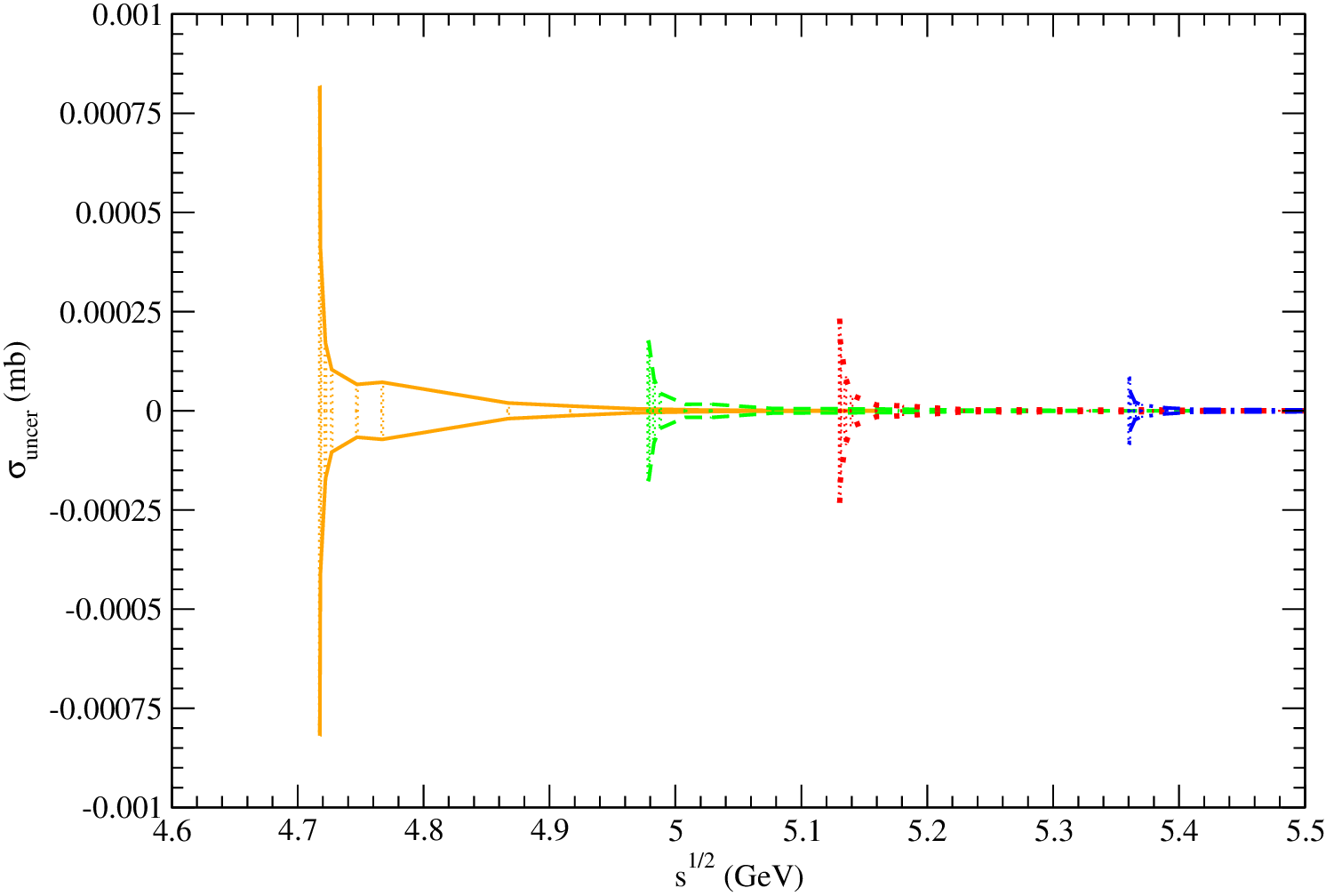}
\caption{The error band between the two solid (dashed, dotted, and dot-dashed)
curves indicates uncertainties of the unpolarized cross sections
for $p \psi(3770) \to \Sigma_c^{++} D^-$
($p \psi(4040) \to \Sigma_c^{++} D^-$, 
$p \psi(4160) \to \Sigma_c^{++} D^-$, and 
$p \psi(4415) \to \Sigma_c^{++} D^-$).}
\label{fig15}
\end{figure}

\newpage
\begin{figure}[htbp]
  \centering
    \includegraphics[scale=0.65]{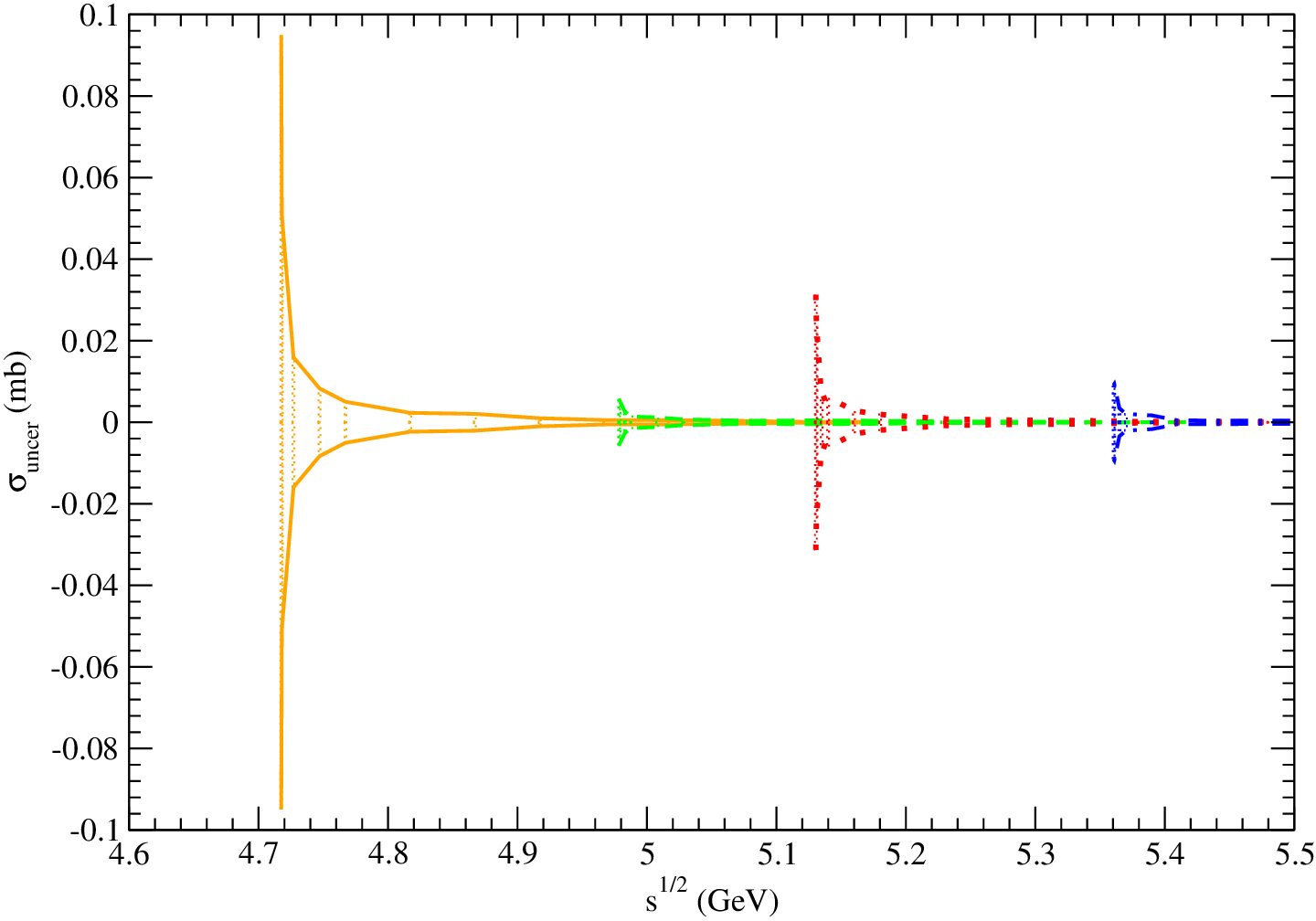}
\caption{The error band between the two solid (dashed, dotted, and dot-dashed)
curves indicates uncertainties of the unpolarized cross sections
for $p \psi(3770) \to \Sigma_c^{++} D^{*-}$
($p \psi(4040) \to \Sigma_c^{++} D^{*-}$, 
$p \psi(4160) \to \Sigma_c^{++} D^{*-}$, and 
$p \psi(4415) \to \Sigma_c^{++} D^{*-}$).}
\label{fig16}
\end{figure}

\newpage
\begin{figure}[htbp]
  \centering
    \includegraphics[scale=0.64]{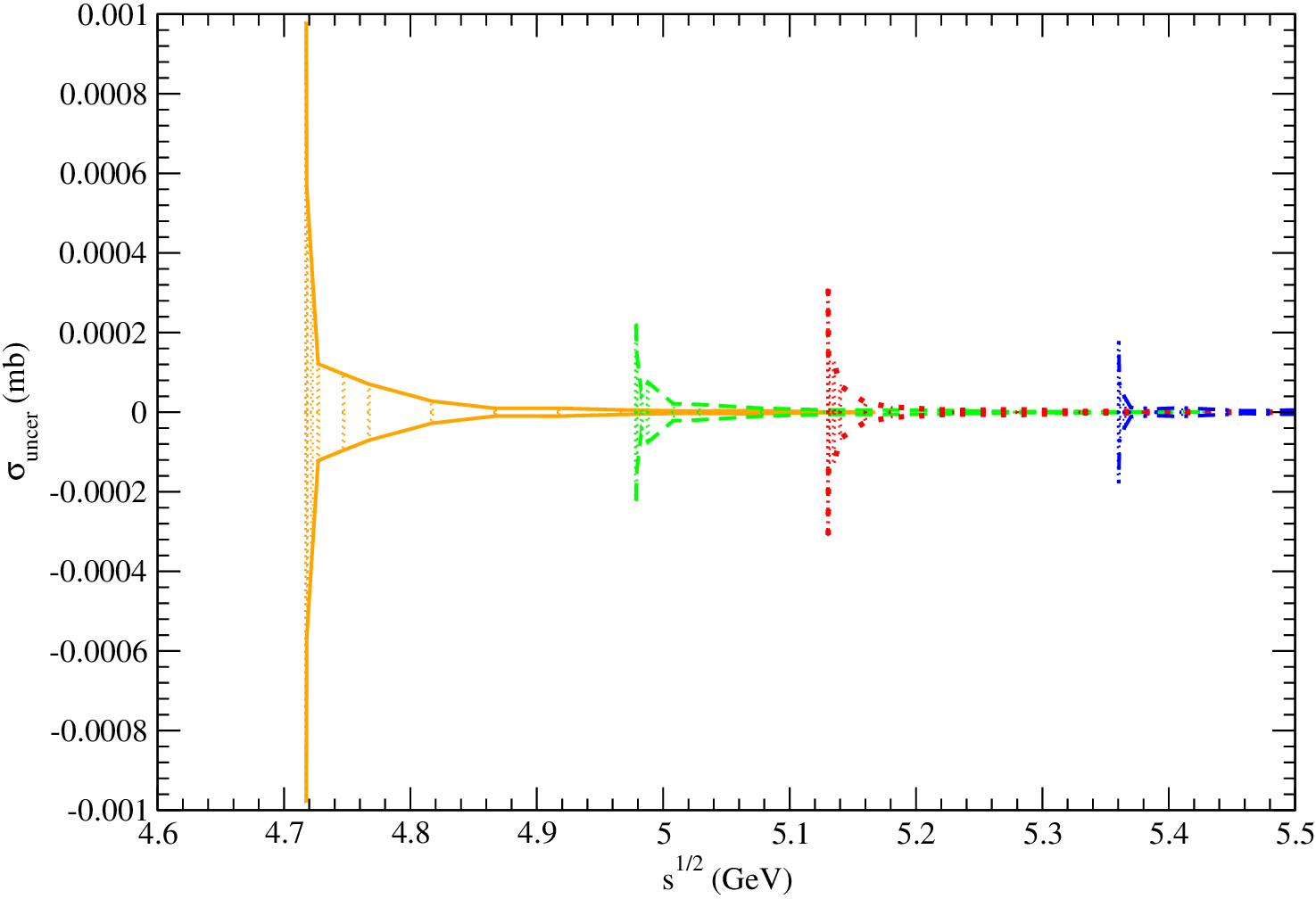}
\caption{The error band between the two solid (dashed, dotted, and dot-dashed)
curves indicates uncertainties of the unpolarized cross sections
for $p \psi(3770) \to \Sigma_c^{+} \bar{D}^0$
($p \psi(4040) \to \Sigma_c^{+} \bar{D}^0$, 
$p \psi(4160) \to \Sigma_c^{+} \bar{D}^0$, and 
$p \psi(4415) \to \Sigma_c^{+} \bar{D}^0$).}
\label{fig17}
\end{figure}

\newpage
\begin{figure}[htbp]
  \centering
    \includegraphics[scale=0.65]{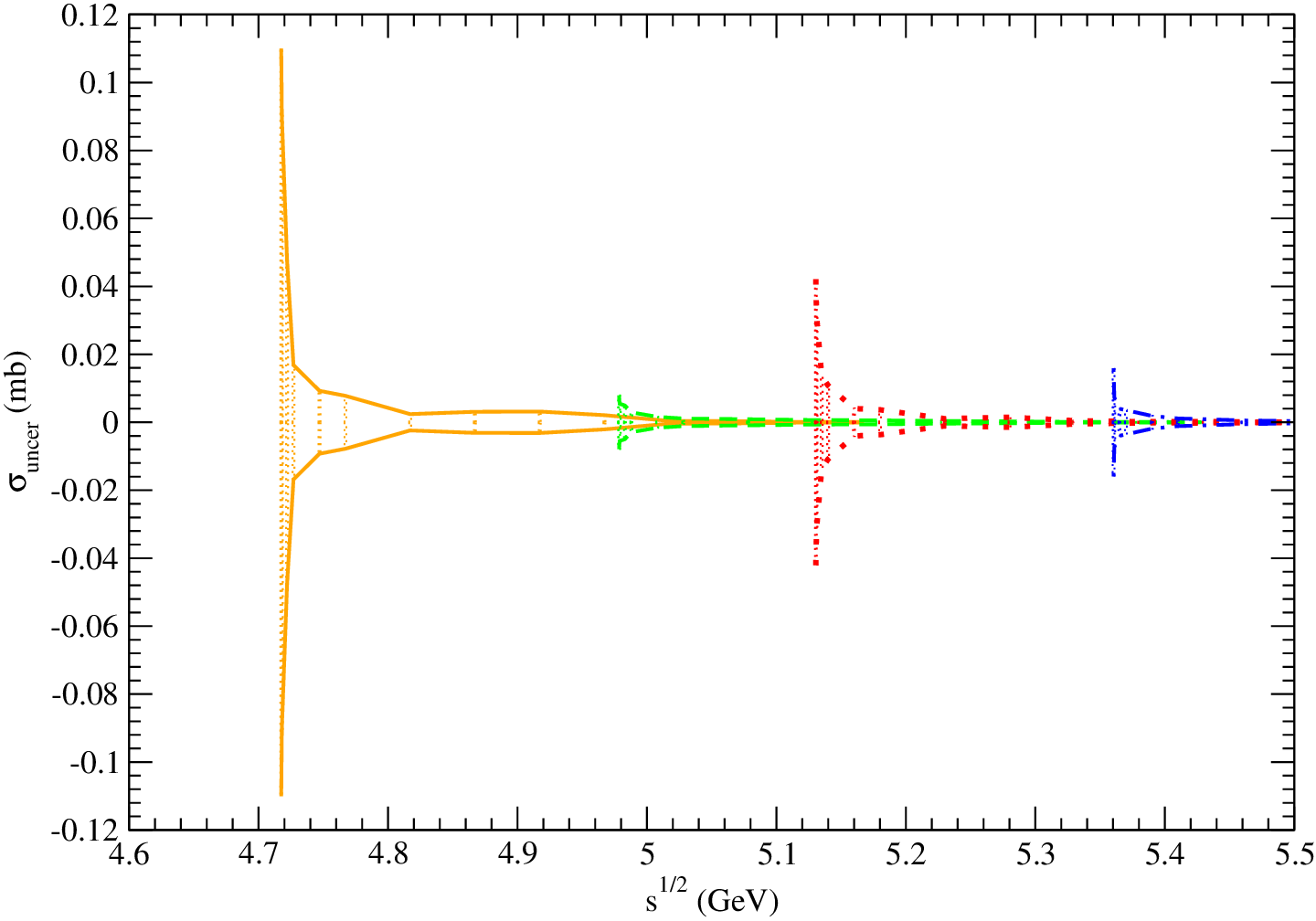}
\caption{The error band between the two solid (dashed, dotted, and dot-dashed)
curves indicates uncertainties of the unpolarized cross sections
for $p \psi(3770) \to \Sigma_c^{+} \bar{D}^{*0}$
($p \psi(4040) \to \Sigma_c^{+} \bar{D}^{*0}$, 
$p \psi(4160) \to \Sigma_c^{+} \bar{D}^{*0}$, and 
$p \psi(4415) \to \Sigma_c^{+} \bar{D}^{*0}$).}
\label{fig18}
\end{figure}

\newpage
\begin{figure}[htbp]
  \centering
    \includegraphics[scale=0.64]{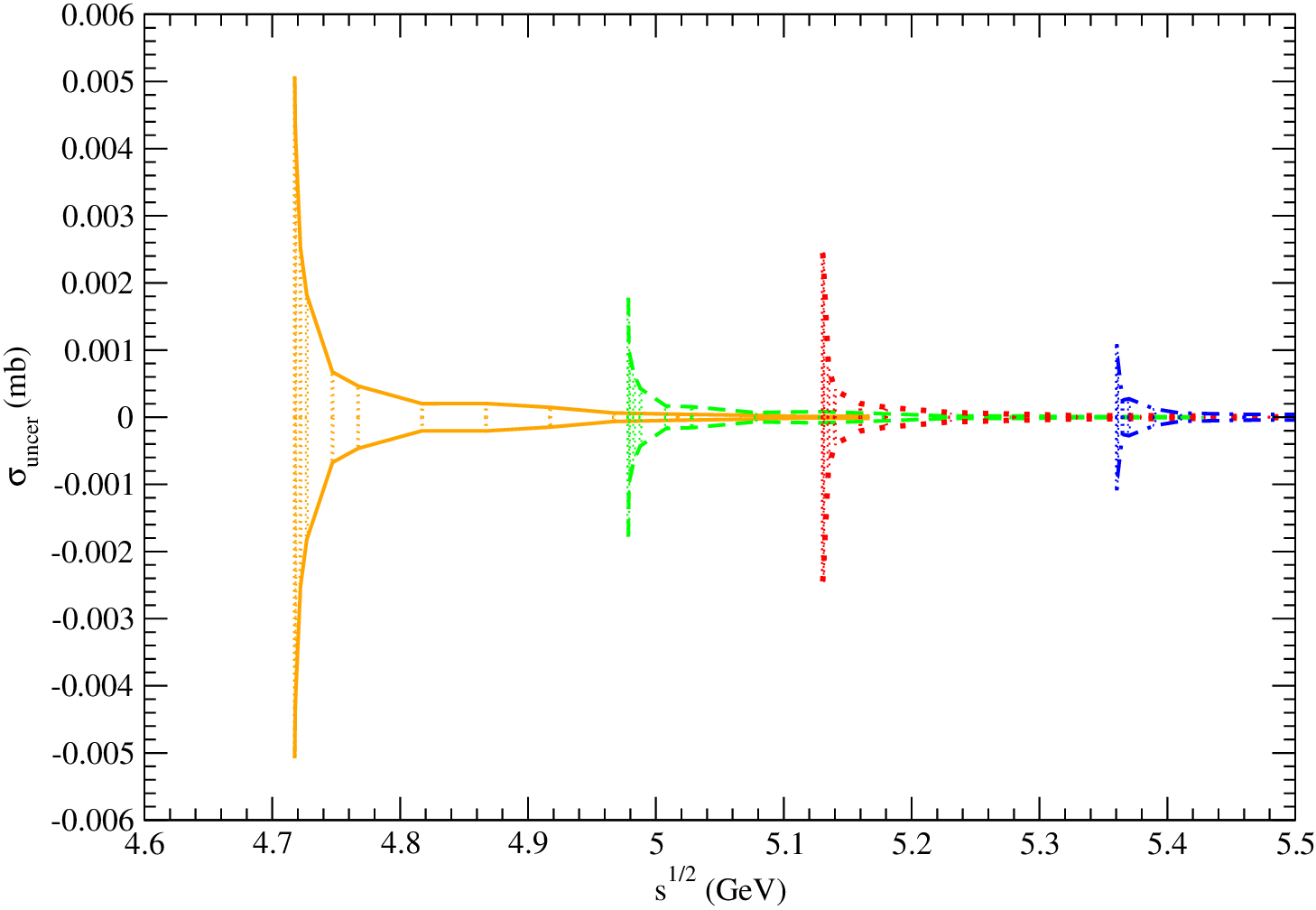}
\caption{The error band between the two solid (dashed, dotted, and dot-dashed)
curves indicates uncertainties of the unpolarized cross sections
for $p \psi(3770) \to \Sigma_c^{*++} D^-$
($p \psi(4040) \to \Sigma_c^{*++} D^-$, 
$p \psi(4160) \to \Sigma_c^{*++} D^-$, and 
$p \psi(4415) \to \Sigma_c^{*++} D^-$).}
\label{fig19}
\end{figure}

\newpage
\begin{figure}[htbp]
  \centering
    \includegraphics[scale=0.64]{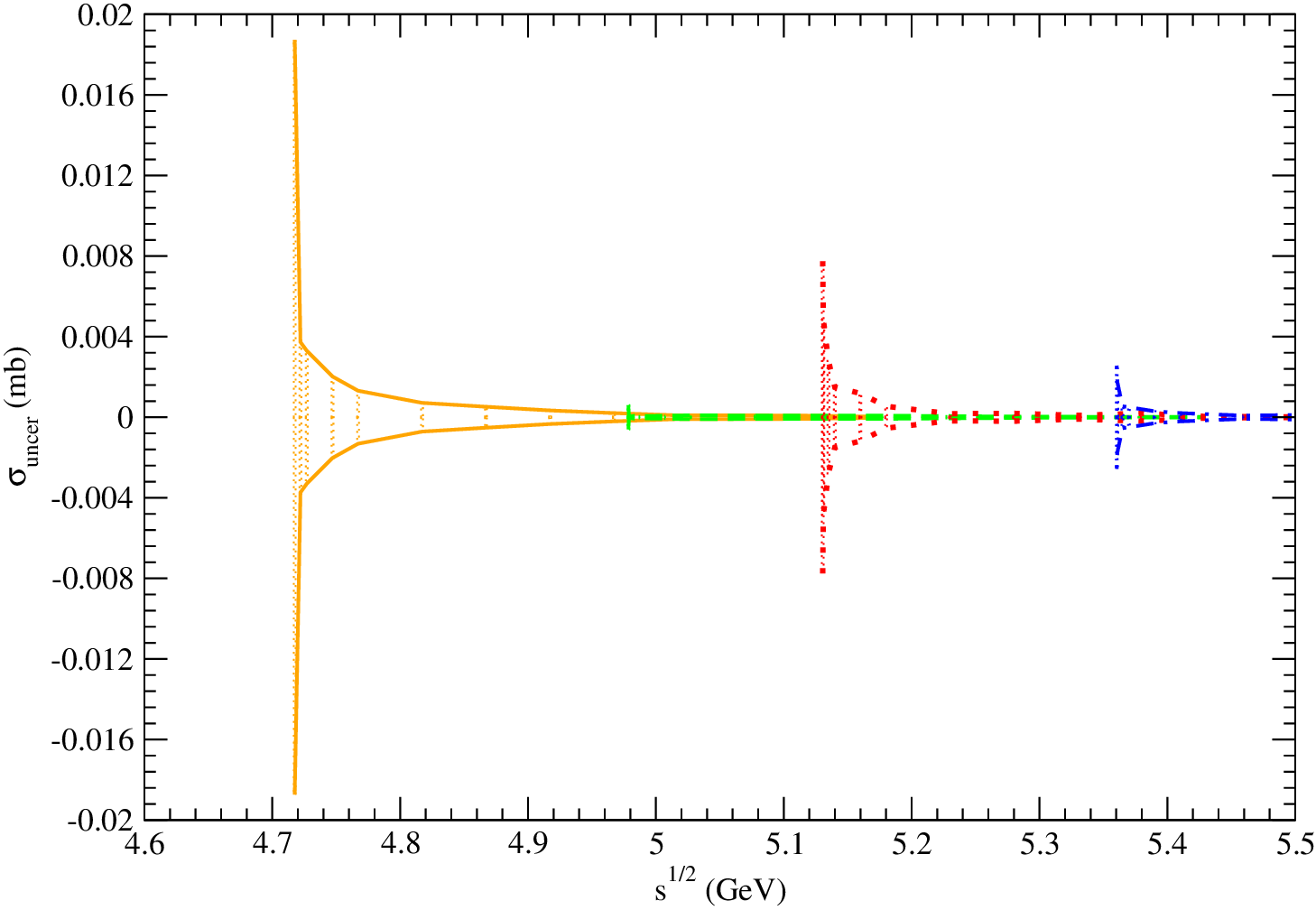}
\caption{The error band between the two solid (dashed, dotted, and dot-dashed)
curves indicates uncertainties of the unpolarized cross sections
for $p \psi(3770) \to \Sigma_c^{*++} D^{*-}$
($p \psi(4040) \to \Sigma_c^{*++} D^{*-}$, 
$p \psi(4160) \to \Sigma_c^{*++} D^{*-}$, and 
$p \psi(4415) \to \Sigma_c^{*++} D^{*-}$).}
\label{fig20}
\end{figure}

\newpage
\begin{figure}[htbp]
  \centering
    \includegraphics[scale=0.65]{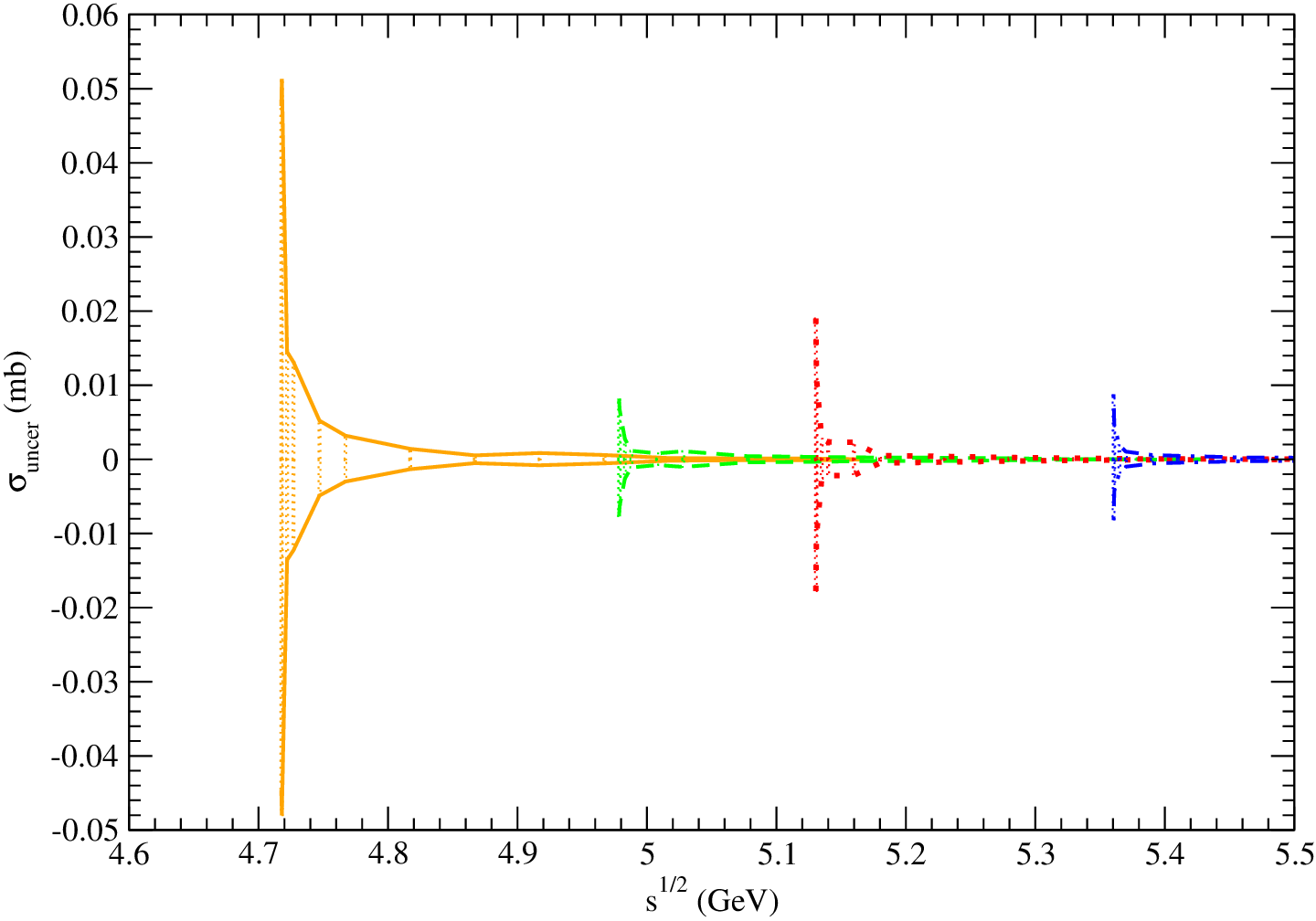}
\caption{The error band between the two solid (dashed, dotted, and dot-dashed)
curves indicates uncertainties of the unpolarized cross sections
for $p \psi(3770) \to \Sigma_c^{*+} \bar{D}^0$
($p \psi(4040) \to \Sigma_c^{*+} \bar{D}^0$, 
$p \psi(4160) \to \Sigma_c^{*+} \bar{D}^0$, and 
$p \psi(4415) \to \Sigma_c^{*+} \bar{D}^0$).}
\label{fig21}
\end{figure}

\newpage
\begin{figure}[htbp]
  \centering
    \includegraphics[scale=0.65]{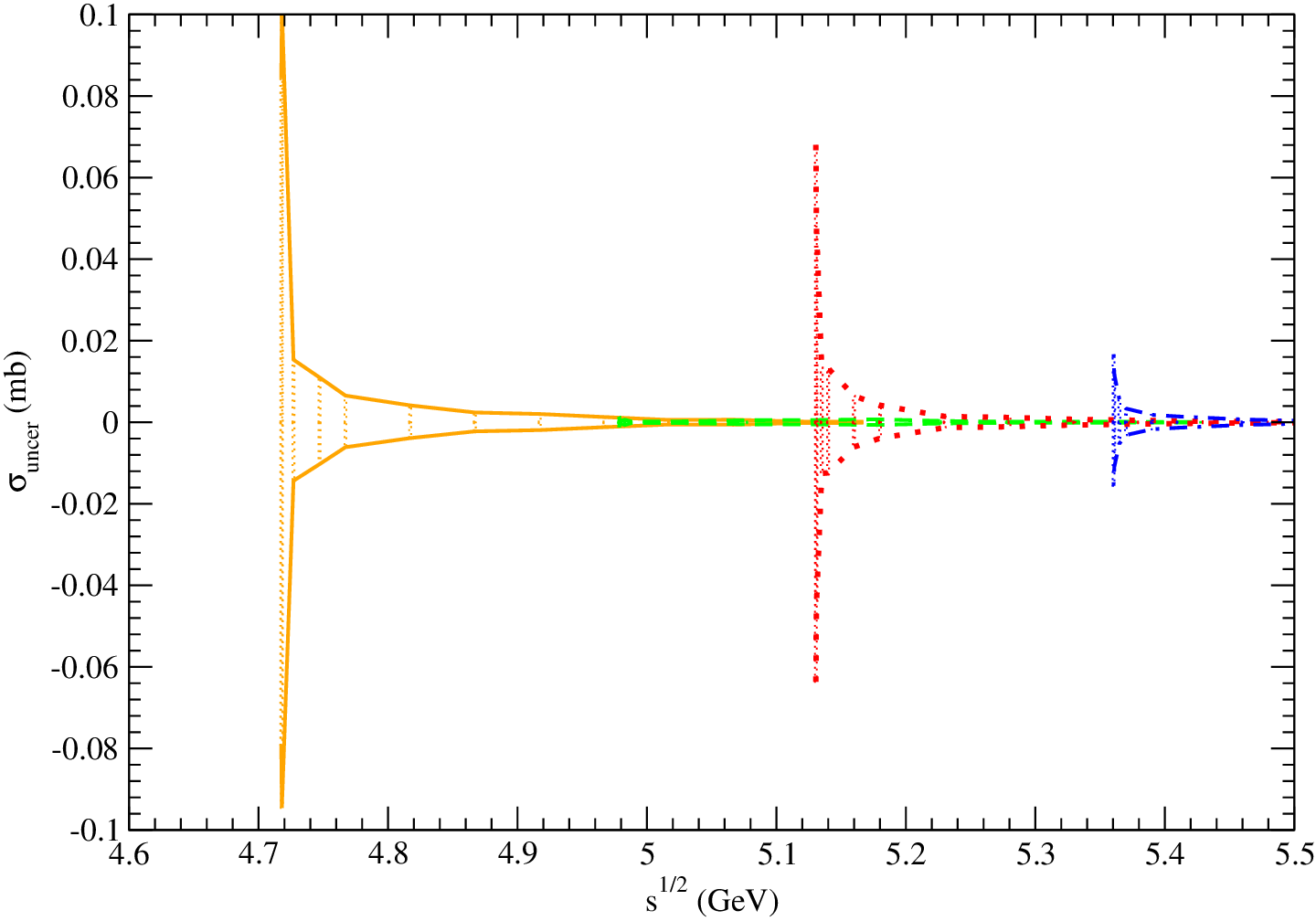}
\caption{The error band between the two solid (dashed, dotted, and dot-dashed)
curves indicates uncertainties of the unpolarized cross sections
for $p \psi(3770) \to \Sigma_c^{*+} \bar{D}^{*0}$
($p \psi(4040) \to \Sigma_c^{*+} \bar{D}^{*0}$, 
$p \psi(4160) \to \Sigma_c^{*+} \bar{D}^{*0}$, and 
$p \psi(4415) \to \Sigma_c^{*+} \bar{D}^{*0}$).}
\label{fig22}
\end{figure}

\newpage
\begin{table*}[htbp]
\caption{\label{table1}Values of the parameters. $a_1$ and $a_2$ are
in units of millibarns; $b_1$ and $b_2$ are in units of GeV; $c_1$ and $c_2$
are dimensionless.}
\tabcolsep=5pt
\begin{tabular}{ccccccc}
  \hline
  \hline
reaction & $a_1$ & $b_1$ & $c_1$ & $a_2$ & $b_2$ & $c_2$ \\
\hline
 $p\psi(3770)\to \Lambda_c^+ \bar{D}^0$
  & 0.13  & 0.01  & 0.51  & 0.23   & 0.09  & 1.08  \\
 $p\psi(4040)\to \Lambda_c^+ \bar{D}^0$
  & 0.065  & 0.068  & 0.41  & 0.017  & 0.18  & 62.4  \\
 $p\psi(4160)\to \Lambda_c^+ \bar{D}^0$
  & 0.026  & 0.032  & 0.55  & 0.018  & 0.18  & 4.67  \\
 $p\psi(4415)\to \Lambda_c^+ \bar{D}^0$
  & 0.0164  & 0.08  & 0.42  & 0.005  & 0.07  & 11.7  \\
 $p\psi(3770)\to \Lambda_c^+ \bar{D}^{*0}$
  & 6.4  & 0.03  & 0.54  & 3.18  & 0.18  & 4.15  \\
 $p\psi(4040)\to \Lambda_c^+ \bar{D}^{*0}$
  & 0.14  & 0.31  & 1.15  & 1.12  & 0.06  & 0.47  \\
 $p\psi(4160)\to \Lambda_c^+ \bar{D}^{*0}$
  & 0.58  & 0.02  & 0.54  & 0.66  & 0.12  & 2.35  \\
 $p\psi(4415)\to \Lambda_c^+ \bar{D}^{*0}$
  & 0.034  & 0.02  & 0.08  & 0.3  & 0.08  & 0.65  \\
 $p\psi(3770)\to \Sigma_c^{++} D^-$
  & 0.017  & 0.07  & 0.93  & 0.021  & 0.05  & 0.35  \\
 $p\psi(4040)\to \Sigma_c^{++} D^-$
  & 0.0067  & 0.071  & 0.51  & 0.004  & 0.19  & 38.6   \\
 $p\psi(4160)\to \Sigma_c^{++} D^-$
  & 0.0012  & 0.01  & 0.35  & 0.0042  & 0.08  & 0.66  \\
 $p\psi(4415)\to \Sigma_c^{++} D^-$
  & 0.0013  & 0.024  & 0.58  & 0.0022  & 0.168  & 4.29  \\
 $p\psi(3770)\to \Sigma_c^{++} D^{*-}$
  & 0.012  & 0.04  & 0.01  & 5.4  & 0.058  & 0.48  \\
 $p\psi(4040)\to \Sigma_c^{++} D^{*-}$
  & 0.34  & 0.03  & 0.64  & 0.53  & 0.19  & 2.33  \\
 $p\psi(4160)\to \Sigma_c^{++} D^{*-}$
  & 0.13  & 0.017  & 5.92  & 0.85  & 0.058  & 0.5  \\
 $p\psi(4415)\to \Sigma_c^{++} D^{*-}$
  & 0.007  & 0.0025  & 0.01  & 0.27  & 0.079  & 0.6  \\
 $p\psi(3770)\to \Sigma_c^+ \bar{D}^0$
  & 0.005  & 0.1  & 0.38  & 0.014  & 0.04  & 0.47  \\
 $p\psi(4040)\to \Sigma_c^+ \bar{D}^0$
  & 0.0009  & 0.01  & 0.29  & 0.0033  & 0.12  & 1.6  \\
 $p\psi(4160)\to \Sigma_c^+ \bar{D}^0$
  & 0.00112  & 0.16  & 2.11  & 0.002  & 0.03  & 0.49  \\
 $p\psi(4415)\to \Sigma_c^+ \bar{D}^0$
  & 0.0008  & 0.07  & 0.437  & 0.0006  & 0.162  & 9.3  \\
  \hline
  \hline
\end{tabular}
\end{table*}

\newpage
\begin{table*}[htbp]
\caption{\label{table2}The same as Table 1, but for twenty other reactions.}
\tabcolsep=5.3pt
\begin{tabular}{ccccccc}
  \hline
  \hline
  reaction & $a_1$ & $b_1$ & $c_1$ & $a_2$ & $b_2$ & $c_2$ \\
  \hline
 $p\psi(3770)\to \Sigma_c^+ \bar{D}^{*0}$
  & 1.4  & 0.01  & 0.52  & 2.6  & 0.1  & 1.21  \\
 $p\psi(4040)\to \Sigma_c^+ \bar{D}^{*0}$
  & 0.041  & 0.0035  & 0.53  & 0.39  & 0.117  & 1.16  \\
 $p\psi(4160)\to \Sigma_c^+ \bar{D}^{*0}$
  & 0.14  & 0.008  & 0.53  & 0.42  & 0.072  & 0.82  \\
 $p\psi(4415)\to \Sigma_c^+ \bar{D}^{*0}$
  & 0.025  & 0.008  & 0.46  & 0.131  & 0.1  & 0.66  \\
 $p\psi(3770)\to \Sigma_c^{*++} D^-$
  & 0.06  & 0.03  & 0.85  & 0.29  & 0.06  & 0.43  \\
 $p\psi(4040)\to \Sigma_c^{*++} D^-$
  & 0.026  & 0.027  & 0.55  & 0.044  & 0.17  & 2.86  \\
 $p\psi(4160)\to \Sigma_c^{*++} D^-$
  & 0.0446  & 0.054  & 0.45  & 0.003  & 0.234  & 82.8  \\
 $p\psi(4415)\to \Sigma_c^{*++} D^-$
  & 0.0037  & 0.24  & 0.21  & 0.013  & 0.07  & 0.63  \\
 $p\psi(3770)\to \Sigma_c^{*++} D^{*-}$
  & 0.4  & 0.09  & 3.24  & 1  & 0.06  & 0.4  \\
 $p\psi(4040)\to \Sigma_c^{*++} D^{*-}$
  & 0.0058  & 0.0038  & 0.38  & 0.0967  & 0.158  & 1.34  \\
 $p\psi(4160)\to \Sigma_c^{*++} D^{*-}$
  & 0.05  & 0.02  & 0.61  & 0.17  & 0.08  & 0.52  \\
 $p\psi(4415)\to \Sigma_c^{*++} D^{*-}$
  & 0.034  & 0.059  & 2.62  & 0.038  & 0.29  & 0.32  \\
 $p\psi(3770)\to \Sigma_c^{*+} \bar{D}^0$
  & 0.005  & 0.013  & 9.57  & 0.178  & 0.059  & 0.49  \\
 $p\psi(4040)\to \Sigma_c^{*+} \bar{D}^0$
  & 0.012  & 0.212  & 8.38  & 0.021  & 0.066  & 0.45  \\
 $p\psi(4160)\to \Sigma_c^{*+} \bar{D}^0$
  & 0.0073  & 0.0161  & 1.5  & 0.019  & 0.061  & 0.4  \\
 $p\psi(4415)\to \Sigma_c^{*+} \bar{D}^0$
  & 0.0002  & 0.001  & 0.5  & 0.0082  & 0.08  & 0.57  \\
 $p\psi(3770)\to \Sigma_c^{*+} \bar{D}^{*0}$
  & 0.148  & 0.186  & 4.97  & 0.583  & 0.0477  & 0.514  \\
 $p\psi(4040)\to \Sigma_c^{*+} \bar{D}^{*0}$
  & 0.0057  & 0.008  & 0.55  & 0.065  & 0.164  & 2.08  \\
 $p\psi(4160)\to \Sigma_c^{*+} \bar{D}^{*0}$
  & 0.037  & 0.013  & 0.33  & 0.1  & 0.09  & 1.25  \\
 $p\psi(4415)\to \Sigma_c^{*+} \bar{D}^{*0}$
  & 0.003  & 0.014  & 0.18  & 0.026  & 0.1  & 0.71  \\
  \hline
  \hline
\end{tabular}
\end{table*}

\end{document}